\documentclass[
reprint,
amsmath,amssymb,
aps,
nofootinbib,
]{revtex4-2}

\usepackage{amsthm}
\usepackage{overpic}
\usepackage{graphicx}
\usepackage{float}
\usepackage{dcolumn}
\usepackage{bm}
\usepackage{adjustbox}
\usepackage{rotating}
\usepackage{ulem}
\usepackage{mathrsfs}
\usepackage{xcolor}
\usepackage{subcaption}
\usepackage{mwe}

\usepackage{algorithm,algcompatible,algorithmicx}
\usepackage{algpseudocode}

\algnewcommand\algorithmicinput{\textbf{Input:}}
\algnewcommand\INPUT{\item[\algorithmicinput]}
\algnewcommand\algorithmicoutput{\textbf{Output:}}
\algnewcommand\OUTPUT{\item[\algorithmicoutput]}
\algnewcommand\algorithmicoptional{\textbf{Optional:}}
\algnewcommand\OPTIONAL{\item[\algorithmicoptional]}

\begin{document}

\title{Reactor-scale stellarators with force and\\ torque minimized dipole coils}%

\author{Alan A. Kaptanoglu}
 \email{alankaptanoglu@nyu.edu}
 \affiliation{Courant Institute of Mathematical Sciences, New York University, New York, NY, 10012, USA \looseness=-1}
 \affiliation{Department of Mechanical Engineering, University of Washington,
Seattle, WA, 98195, USA \looseness=-1
}
\author{Alexander Wiedman}
\affiliation{IREAP, University of Maryland, College Park, MD, 20742, USA \looseness=-1}
\author{Jacob Halpern}
\affiliation{APAM, Columbia University, New York City, NY, 10027, USA \looseness=-1}
\author{Siena Hurwitz}
\affiliation{IREAP, University of Maryland, College Park, MD, 20742, USA   \looseness=-1}
\author{Elizabeth J. Paul}
\affiliation{APAM, Columbia University, New York City, NY, 10027, USA \looseness=-1}
\author{Matt Landreman}
\affiliation{IREAP, University of Maryland, College Park, MD, 20742, USA   \looseness=-1}
\begin{abstract}
In this work, we utilize new coil objectives for stellarator optimization with autodifferentiation, including pointwise and net coil-coil forces and torques. We use these methods to perform the first large-scale optimization of planar dipole coil arrays, since arrays of small and geometrically simple coils have been proposed to partially produce the 3D magnetic fields for stellarators, generate advantageous magnetic field perturbations in tokamaks, and provide active, real-time control capabilities. We perform an ablation study to show that minimizing the orientation and location of each coil may be essential to get coil forces, coil torques, and field errors to tolerable levels. We conclude with solutions for three reactor-scale quasi-symmetric stellarators \textcolor{black}{by jointly optimizing nonplanar TF coils and planar coil arrays}.

 \end{abstract}
\maketitle

\section{Introduction}
\label{sec:intro}
Coil design is a critical component of designing stellarators, a class of plasma devices commonly considered for future nuclear fusion reactors based on  plasma confinement from strong magnetic fields. Designing coils is a form of inverse magnetostatics, an ill-posed inverse problem stemming from the reality that many different magnet designs can produce a nearly identical target magnetic field via the Biot-Savart law~\cite{imbert2019introduction}. 
This ill-posedness is often an advantageous feature; the problem has extra degrees of freedom which can be used to impose additional engineering constraints related to the cost and complexity of the coils.  

Optimizing stellarators is often performed in two stages. The first is a shape optimization problem that is solved using fixed-boundary magnetohydrodynamic (MHD) equilibrium codes to find plasma equilibria with desirable physics properties~\cite{nuhrenberg1988quasi,spong1998j,drevlak2018optimisation,landreman2021simsopt,dudt2023desc,landreman2022magnetic,kim2024optimization,hegna2022improving}. 
After obtaining an equilibrium in the first stage, coil optimization is performed to determine a set of magnets that can produce a magnetic field that matches the boundary conditions at the assumed plasma surface, subject to engineering constraints such as minimum coil-coil distances, maximum forces on the coils, maximum curvature on the coils, and many other requirements. 
Unfortunately, stellarator coils are often geometrically complex, require tight tolerances, and therefore exhibit high construction costs, assembly costs, and maintenance challenges. It is well-known that the W-7X and NCSX stellarator program costs and delays came primarily from difficulties with the coils~\cite{erckmann1997w7,strykowsky2009engineering}. 

There appear to be some alternative options for producing the required magnetic fields for stellarator design that trade the complexity of standard modular coils, i.e. coils that are singly-linked with and fairly close to  the toroidal plasma, with the complexity of solving more difficult or higher-dimensional optimization problems.
For instance, permanent magnets can be effectively used with simple planar coils in laboratory-scale devices~\cite{qian2022simpler,hammond2020geometric,kaptanoglu2022greedy,kaptanoglu2022permanent,hammond2024improved,yu2024quasi}, but thousands or tens of thousands of magnets may be required, the magnets have limited field strengths, and they demagnetize in the high-field environments relevant for nuclear fusion reactors. Nonetheless, the MUSE experiment~\cite{qian2023design} illustrated that a laboratory-scale stellarator could be built very cost-effectively by using such a modular source array. Others, such as the start-up companies Thea Energy~\cite{gates2023thea} and Stellarex~\cite{zarnstorff2023stellarex}, have proposed large arrays of relatively small saddle/dipole coils (coils that are not topologically linked with the plasma) to provide the benefits of permanent magnets (simplicity, modularity, cost, etc.) without the primary downsides (limited field strengths, cannot turn off the fields). A small number of large saddle coils or dipole coils have previously been considered for NCSX~\cite{williamson1999design} and CSX~\cite{baillod2024integrating}. \textcolor{black}{In addition, dipole arrays can have dynamically controlled currents for fixing error fields from unmodelled magnetic field contributions, e.g. from small coil misalignments, or for responding to time-dependent plasma dynamics.}

Such arrays cannot produce a net toroidal magnetic flux because they do not link the torus~\cite{helander2020stellarators}, so for all arrays, a simple set of toroidal field (TF) coils are still required. Nonetheless, the reduction in cost and complexity of the final coil design may be significant when using coil arrays. For simplicity in design, optimization, and construction, we will assume throughout the present work that each dipole coil is a \textit{planar} coil with small cross section. This has the advantage of geometric simplicity and facilitates the direct optimization of the orientation of each coil (rather than a set of Fourier coefficients parametrizing a 3D curve, for which directly optimizing the orientations is cumbersome). 

However, it remains to be shown definitively in the literature that a reactor-scale stellarator can be designed with dipole arrays.
It is not a-priori clear that designing reactor-scale dipole array stellarators is feasible; in order to compensate for the small size of each dipole coil and $r^{-3}$ falloff of a dipole magnetic field, it seems plausible that a reactor-scale design would demand a large number of coils, unreasonably large currents, and/or coils placed close to the plasma, in order to provide the required field strengths. These large currents inevitably lead to large forces, unless additional degrees of freedom can be introduced to mitigate them. Moreover, reactor-scale coils must be approximately $1.5$m away from the plasma or further in order to facilitate a neutron-absorbing blanket.

\subsection{Contributions of this work}
In this work, we provide a first demonstration of large-scale optimization of planar coil arrays and illustrate that such arrays with simple TF coils can be effectively used for reactor-scale stellarators. In order to control for large forces and torques on the dipole coils, we implement new coil objectives using autodifferentiation tools through JAX~\cite{bradbury2018jax} and show that letting the dipole coils move and rotate in space can be a very effective way to reduce forces and torques.
To conclude, coil solutions with a minimum $1.5$ meter coil-plasma distance and other requirements are shown for three reactor-scaled vacuum stellarators. Through an ablation study, we show that force and torque minimization can be essential for designing dipole arrays, and net torques on the TF coils can be especially reduced. The entirety of this work is implemented and can be reproduced in the open-source SIMSOPT code~\cite{landreman2021simsopt}. The configuration files to generate the figures and results in the present work can be found at  \url{https://doi.org/10.5281/zenodo.14934092}.

\textcolor{black}{
\subsubsection*{Comparison with recent work}}

\textcolor{black}{Recent work from Thea energy~\cite{Wu_2025,gates2025stellarator,kruger2025coil} indicate that modular dipole coil array solutions can be found for a DD fusion neutron-source design, even with the constraint that the TF coils are planar, but with somewhat larger normalized field errors than in the present work. For instance, $\langle \bm B\cdot\hat{\bm n}\rangle/\langle B \rangle \approx 0.00358$ is achieved in~\cite{kruger2025coil} while the least accurate solution in our summary Table~\ref{tab:dipole_arrays_summary} in Sec.~\ref{sec:results} is $\langle \bm B\cdot\hat{\bm n}\rangle/\langle B \rangle \approx 0.0019$. It is unclear in~\cite{kruger2025coil} whether this level of average field errors is sufficient to avoid degradation of Poincaré plots or the desired quasi-symmetry. 
}

\textcolor{black}{
Moreover, the use of DD fusion facilitates a thinner neutron-absorbing blanket, so that the TF and dipole coils can be half the distance from the plasma ($\sim 75$ cm instead of $\sim 150$ cm in this work). The consequence is that dipole fields are made slightly more effective, since the approximate $r^{-1}$ falloff of the fields, after rescaling the dipole sizes as $r^2$, means that the magnetic fields from dipoles at $75$ cm distances are a factor of two stronger than ones at $1.5$ m. This factor of two does not fully explain why their dipole solutions, scaled to reactor-size, have $\sim 2-3$ MA of current and our solutions usually exhibit closer to $\sim 10$ MA of current. As we make more clear below, the remaining difference in the maximum currents likely comes (1) the higher field accuracy of our solutions, and (2) the other work using many more dipoles, although it is hard to say because of the significant geometrical differences between these designs. 
In the scenarios we considered, our results illustrated (1) the importance of directly minimizing forces and torques in the reactor-scale setting where currents can be very large, and (2) the efficiency of a reactor-scale dipole array stellarator improves if a thin neutron blanket can be used (e.g. because of material science advances, the use of DD fusion, 3D deposition asymmetries, and so on). For instance, the recent Type One Energy DT reactor design plans for a neutron-absorbing blanket as thin as $\sim 0.8$m in certain locations~\cite{hegna2025infinity}.
}

\textcolor{black}{
The Thea Energy design additionally uses a very different optimization scheme than in the present work. A separated two-stage approach is employed in which the TF coils are first optimized alone, and afterwards the dipole array currents are optimized. Dipole array orientations, locations, and shapes are all fixed. In contrast, we have performed joint optimization of the TF and dipole coils, and can fully vary the dipole currents, orientations, spatial locations, and shapes. We show substantial evidence in the present work that joint optimization and these extra degrees of freedom can be useful for finding a solution in the reactor-scale regime.}

Finally, the Thea Energy final design employs a total of 712 dipole coils in their final solution, while at most we use 216 and at fewest use only 64. However, their coils lie on a single toroidal winding surface and their TF coils are planar, greatly improving the ease of construction and maintenance of the machine. The optimal tradeoff will depend on what strategies are adopted for cost-effectively building modular support structures for the dipole coil arrays. Moreover, we find that the requirements on the dipole array can be much more stringent for some plasmas. For instance, the Landreman-Paul QA and QH stellarators considered here require very high field accuracy from the coils in order to retain their high levels of quasi-symmetry.

\section{New coil objectives}
The stellarator coil optimization literature has increasingly shown that coil-coil forces should be considered during coil optimization. The three-dimensionality and large currents flowing in TF coils, a few Mega-Amperes (MA) to tens of MA for reactor-scale devices, imply very large coil-coil forces. There has been some recent optimization work with coil forces: Volpe et al.~\cite{robin2022minimization} and Fu et al.~\cite{fu2024global} minimize coil-coil forces with a winding surface formulation, Hurwitz and Landreman ~\cite{hurwitz2024efficient,landreman2023efficient} have derived self-force and self-inductance expressions for finite-thickness circular and rectangular cross-section wires and implemented them in optimization~\cite{hurwitz2024opt}, and Guinchard et al.~\cite{guinchard2024including} implemented a penalty on the total magnetic energy in a filamentary representation. 
In this work, we reproduce the filamentary total energy objective in Guinchard et al. and the pointwise-force terms from Hurwitz et al.~\cite{hurwitz2024efficient}, and implement additional net force and torque calculations using autodifferentiation. Joint minimization of net and pointwise forces and torques is important, as they address differing engineering issues.

\section{Planar and nonplanar coil representations}\label{sec:coil_representations}
Planar coils are advantageous for use in large coil arrays because of the geometric simplicity and  the subsequent possibility of cheap mass production and assembly. Another advantage is the ability to directly optimize coil orientation degrees of freedom (DOFs). The SIMSOPT code allows for curves with different representations to inherent from a JAX-based JaxCurve object, that effectively auto-differentiates any function of the curve DOFs. Autodifferentiation is increasingly used for stellarator optimization~\cite{mcgreivy2021optimized,dudt2023desc}. All that is required to optimize any objectives that depend on the curve DOFs is to specify the DOFs and the transformation from those DOFs to the Cartesian coordinates of the curve. However, we found in practice the speed of the code is substantially improved if we used a curve representation written in parallelized C++, and only use JAX for computing derivatives of the optimization objectives with respect to the coil position vectors.

The following finite Fourier series representation of order $M$ is used for representing the polar coordinate $(r, \phi)$ of a planar coil, measured from the center, which gives the location of a point on the coil:
\begin{align}
    r(\phi) = \sum_{m=0}^M r_{c, m}\cos(m\phi) + \sum_{m=1}^M r_{s, m}\sin(m \phi).
\end{align}
To convert this to a representation in a global coordinate system~\cite{wiedman2023coil}, define the coil center at $(X, Y, Z)$, and rotate the coil by a normalized quaternion $[a, b, c, d]$,
\medmuskip=0mu
\thinmuskip=0mu
\thickmuskip=0mu
\begin{align}
\notag
    x &= (1 - 2(c^2 + d^2))r(\phi)\cos(\phi) + 2(bc -ad))r(\phi)\sin(\phi) + X, \\
    \notag
    y &= 2(ad + bc))r(\phi)\cos(\phi) + (1 - 2(b^2 + d^2))r(\phi)\sin(\phi) + Y, \\
    z &= 2(bd - ac))r(\phi)\cos(\phi) + 2(ab + cd))r(\phi)\sin(\phi) + Z.
\end{align}
\medmuskip=2mu
\thinmuskip=2mu
\thickmuskip=2mu
The use of quaternions prevents gimbal lock and other issues with using a representation based on Euler angles. Along with the current in the coil, it follows that the degrees of freedom for each coil are \begin{align}
    \bm \eta = [r_{c,0},...,r_{c,M},r_{s,1},...,r_{s,M},a',b',c',d',X,Y,Z, I],
\end{align}
where $[a', b', c', d']$ represents an unnormalized quaternion, and altogether the DOFs are $2M + 9$ real numbers. If the array of planar coils consists of $N$ unique coils, there are $N(2M + 9)$ degrees of freedom. However, in this work we are interested in fixed and identical circular curves, $M = 0$ and $r_{c,0}$ constant, so there are only $8N$ degrees of freedom. \textcolor{black}{Nonetheless, we emphasize that future work could immediately extend some shape degrees of freedom to the coils while maintaining the planarity, as in~\cite{kaptanoglu2025optimization}. For instance, an array of elliptical planar coils with varying major and minor radii could still be conducive to mass production while facilitating higher solution accuracy.}

Note that in general there are rotated and reflected copies of each unique coil so that the overall arrangement produces magnetic fields that are $n_{p}$-field-period symmetric and stellarator symmetric, which is characteristic of most stellarator designs. Therefore, it is only required to design the coils in the $\zeta \in [0, \pi/n_p)$ sector. Here $\zeta$ is the usual toroidal angle, to distinguish it from $\phi$ which denotes the parametrization of a coil curve. The remaining coils outside the unique sector have no degrees of freedom; they are obtained by applying the symmetries.

While it would be desirable to represent the TF coils in this work by planar coils,
in practice we found, as in Wiedman et al.~\cite{wiedman2023coil}, that the restriction of planar TF coils often prevents adequate accuracy in the final coil set. The restriction of planar TF coils tends to put additional requirements on the dipole coil fields, which are already heavily constrained by the force and torque tolerances. Nonetheless, we show in Sec.~\ref{sec:results} that TF coil complexity can be substantially reduced by using dipole coil arrays. 

Therefore, we consider nonplanar TF coils with the traditional representation as a 3D curve in Cartesian space defined by a finite Fourier series with maximum Fourier mode number $M'$~\cite{zhu2017new},
\textcolor{black}{
\begin{align}
     \bm x(\phi) &= \sum_{m=0}^{M'} \bm x_{c, m}\cos(m\phi) + \sum_{m=1}^{M'} \bm x_{s, m}\sin(m \phi),
     \\ \notag
     \bm \eta' &= [x_{c,0}, x_{s, 1}, ...,x_{c,M'},y_{c, 0},...,y_{c,M'},z_{c, 0},..,z_{c,M'}, I].
\end{align}
With the coil current included, the vector $\bm \eta'$ represents $3N'(2M' + 2)$ degrees of freedom, for a total of $N_{\eta+\eta'} = 8N + 3N'(2M' + 2)$. A typical result in this work uses, $N\sim 30$, $N' \sim 3$, $M' \sim 4$, for a rough total of 330 degrees of freedom, a moderate-dimensional optimization problem. Note that if we fix the coil locations and orientations, as we do later in this work, $N_{\eta+\eta'} = N + 3N'(2M' + 2) \sim 120$ degrees of freedom.} We will use $\sim$ to indicate a factor of two or so throughout this paper. 

\section{Optimization}\label{sec:optimization_problem}
Joint optimization of a set of very different coils, here a small set of large TF coils and large set of small dipole coils, requires some objective terms that only penalize one set or the other. Denote the plasma surface $S$.
We formulate the following optimization problem using well-known objectives for penalizing the max and mean-square coil curvatures, the coil-coil linking numbers, and the total coil lengths. We will denote the $i$-th coil curve by $C_i$, the current by $I_i$, the curvature by $\kappa_i$, the Gauss linking number integral formula between coils $i$ and $j$ as $Lk_{ij}$, the pointwise coil-coil force per length as $d\bm F_i/dl_i$, the pointwise coil-coil torque per length as $d\bm \tau_i/dl_i$, the minimum distance between the $i$-th coil and the plasma surface as $d_{cs, i}$, the minimum distance between coils $i$ and $j$ as $d_{ij}$, and the curve length of coil $i$ as $L_i$. In addition we add various force and torque-related terms and the total vacuum magnetic energy to the optimization problem:
\medmuskip=0mu
\thinmuskip=1mu
\thickmuskip=1mu
\begin{align}
\label{eq:optimization_joint}
    \min_{\bm\eta, \bm \eta'}\left[f_B(\bm \eta, \bm \eta') + f_\text{TF}(\bm \eta') + f_\text{mixed}(\bm \eta, \bm \eta') \right], 
    \end{align}
    \begin{align}
    f_B(\bm \eta, \bm \eta') = &\frac{1}{2}\int_S|\bm B(\bm \eta, \bm \eta ')\cdot\hat{\bm n}|^2dS, 
    \end{align}
    \begin{align}
    \notag
    f_\text{TF}&(\bm \eta') = \frac{w_L}{2}\sum_{i=1}^{N}\max(L_i - L_0, 0)^2 + \frac{w_c}{2}\oint_{C_i}\max(\kappa_i - \kappa_0, 0)^2 dl \\  + &\frac{w_\text{msc}}{2}\sum_{i=1}^N \max\left(\left(\frac{1}{L_i}\oint_{C_i}\kappa_i^2(l_i) dl_i\right) - \text{MSC}_0, 0\right)^2,      \label{eq:TF_penalty}
  \end{align}
  \begin{align}
    &f_\text{mixed}(\bm \eta, \bm \eta') = w_\text{Lk} \sum_{i \neq j}^{N + N'} Lk_{ij} + \frac{w_\text{tve}}{2}\sum_{i,j=1}^{N + N'}I_iL_{ij}I_j\\ \notag + &w_\text{cc}\sum_{i\neq j}^{N + N'}\max\left( d_{cc,0} - d_{ij}, 0\right)^2 +w_\text{cs}\sum_{i=1}^{N + N'}\max\left( d_{cs,0} - d_{cs,i}, 0\right)^2 \\ \notag + &w_\text{nf}\sum_{i=1}^{N + N'} \left|\oint_{C_i}\frac{d\bm F_i(\bm \eta, \bm \eta')}{dl_i}dl_i\right|^2 \\ \notag + &w_\text{nt}\sum_{i=1}^{N + N'} \left|\oint_{C_i}\frac{d\bm \tau_i(\bm \eta, \bm \eta')}{dl_i}dl_i\right|^2 \\ \notag + &w_\text{pf}\sum_{i=1}^{N + N'} \oint_{C_i}dl_i \max\left(
    \left|\frac{d\bm F_i}{dl_i}\right| - \left|\frac{d\bm F}{dl_i}\right|_0, 0\right)^2 \\ \notag + &w_\text{pt}\sum_{i=1}^{N + N'} \oint_{C_i}dl_i \max\left(
    \left|\frac{d\bm \tau_i}{dl_i}\right| - \left|\frac{d\bm \tau}{dl_i}\right|_0, 0\right)^2.
\end{align}
\medmuskip=4mu
\thinmuskip=4mu
\thickmuskip=4mu
Here $w_L$, $w_c$, $w_\text{cc}$, $w_{sc}$, $w_\text{msc}$, $w_\text{Lk}$, $w_\text{nf}$, $w_\text{nt}$, $w_\text{pf}$, $w_\text{pt}$, and $w_\text{tve}$ are hyperparameter weights. 
$\kappa_0$ is the threshold curvature above which the penalty applies, and similarly for  $d_{cc,0}$, $d_{cs,0}$, $L_0$, $ \|d\bm F/dl_i|_0$, $ \|d\bm \tau/dl_i|_0$ and $\text{MSC}_0$. 
Similar objectives have been defined in previous filamentary optimization work, see e.g. Wiedman et al.~\cite{wiedman2023coil} or Wechsung et al.~\cite{wechsung2022precise}. For two filamentary coils, the pointwise force and torque on coil $i$, as well as the inductance matrix $L_{ij}$ between all coils, are computed via appropriate Neumann integral formulas,
\medmuskip=0mu
\thinmuskip=0mu
\thickmuskip=0mu
\begin{align}
    \frac{d\bm F_i}{dl_i} &=  \frac{d\bm F_\text{self}}{dl_i} -\sum_{j\neq i}\frac{\mu_0 I_iI_j}{4\pi}\bm t_i\times \oint_{C_j} \frac{d\bm l_j \times\bm r_{ij}}{|\bm r_{ij}|^3}, \\  
    \frac{d\bm \tau_i}{dl_i} &= (\bm r_i-\bm c_i) \times \frac{d\bm F_i}{dl_i}, \\
    L_{ij} &= \frac{\mu_0}{4\pi}\oint_{C_i}\oint_{C_j} \frac{d\bm l_i\cdot d\bm l_j}{|\bm r_{ij}|}, \quad i \neq j, \\ L_{ii} &= \frac{\mu_0}{4\pi}\oint_{C_i}\oint_{C_i} \frac{d\bm l_i\cdot d\bm l_i}{\sqrt{|\bm r_{ii}|^2 + \delta ab}}, \label{eq:f_mixed}
\end{align}
\medmuskip=4mu
\thinmuskip=4mu
\thickmuskip=4mu
where $\bm t_i$ is the tangent vector to the curve $C_i$, $\bm c_i$ is the barycenter of $C_i$,
\begin{align}
    \bm c_i = \frac{1}{L_i}\int_{C_i}\bm r_i dl_i,
\end{align}
$\bm r_{ij} = \bm r_i - \bm r_j$, and $\delta$, $a$, $b$ are regularization constants that can be found in Hurwitz and Landreman~\cite{hurwitz2024efficient}, along with a fast method to evaluate $L_{ii}$. The self-forces and self-torques are obtained from the expressions in Hurwitz and Landreman~\cite{hurwitz2024efficient}. Note that there are no net self-forces or net self-torques, so we remove these terms for calculating the net forces and torques.

Taken in full, this problem has $11$ objective weights and $7$ threshold values, for a total of $18$ hyperparameters. In practice, this work finds that the results in the present work are insensitive to a number of simplifications: $w_c = w_\text{msc} = \kappa_0 = MSC_0 = w_\text{tve} = 0$. Removing the curvature objectives is justified because the curvature can often be reasonably well-controlled by tuning the order of the Fourier expansions and the penalty on the coil lengths, and for the purposes of dipole array optimization, all the coils in the array will have fixed circular shapes. 
Moreover, large optimization scans performed in Appendix~\ref{sec:appendix_pareto} found strong correlations between the total vacuum energy and the coil forces, the coil torques, and the coil lengths. Similar observations, such as the very strong correlation with total coil length, were also found by Guinchard et al.~\cite{guinchard2024including}. For this reason, we drop the total vacuum energy term for the remainder of this work.
In addition, we often find that pointwise torques and net forces do not need to be directly optimized in the coil configurations investigated in the present work, $w_\text{pt} = \|d\bm\tau_i/dl_i \|_0 = w_\text{nf} = 0$. The pointwise forces are heavily penalized and correlate with the other force and torque objectives, as we show in Sec.~\ref{sec:preliminary_scans}.
Finally, we optimize Eq.~\eqref{eq:optimization_joint} with the standard L-BFGS algorithm~\cite{liu1989limited} available through \textit{scipy.minimize}. 

\section{Preliminary Scans}\label{sec:preliminary_scans}
We begin without dipole arrays to validate some of the new coil objectives in this work. Each of the newly-implemented objectives  are verified against finite differences. For further validation, we reproduce a standard four TF coil per half-field period example in the SIMSOPT code that illustrates basic stage-two optimization for the Landreman and Paul precise-QA design~\cite{landreman2022magnetic} scaled to $1$ m major radius and a modest $0.3$ T magnetic field strength along the magnetic axis. We also assume a single turn of wire, so that the absolute force magnitudes here will appear quite large for a small device. The forces vary inversely with the number of turns of wire.

\begin{figure*}
    \centering
    \includegraphics[width=0.85\linewidth]{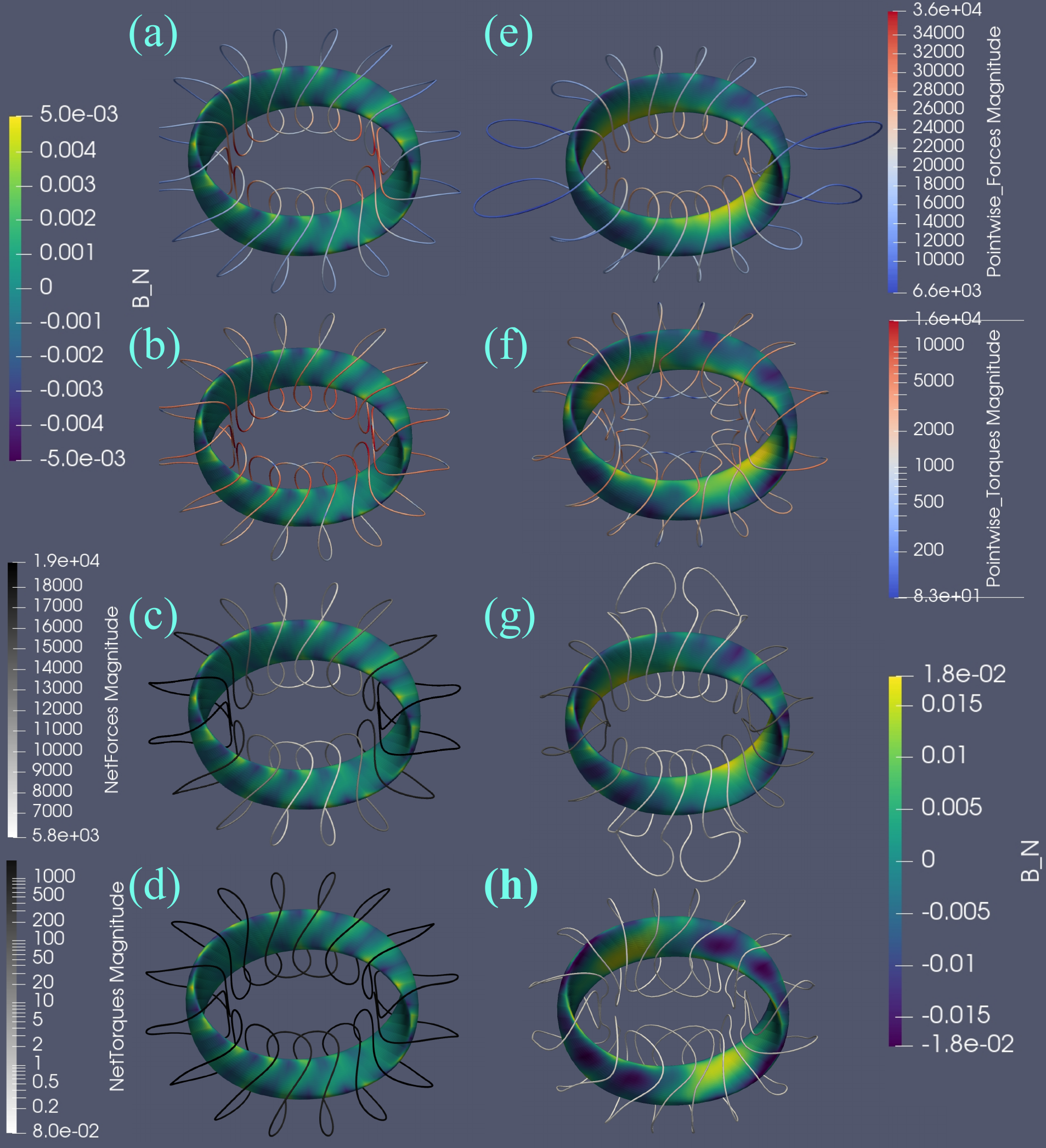}
    \caption{\textcolor{black}{Left column: A single coil set without force and torque optimization, visualizing the coil (a) pointwise forces, (b) pointwise torques, (c) net forces, and (d) net torques. Right column (e)-(h): same as left, but the visualized coil quantity has been directly optimized.} There are separate $\bm B\cdot\hat{\bm n}$ colorbars for the left and right columns.}
    \label{fig:force_scan}
\end{figure*}

We begin by scanning the new objective weights, e.g. $w_\text{nf}$, keeping the other new weights fixed at zero, $w_\text{nt} = w_\text{pt} = w_\text{pf} = 0$. Figure~\ref{fig:force_scan} shows the difference in solution quality and coil design between an optimized solution set when net and pointwise forces and torques are minimized or not minimized. We compare these runs at roughly the same $\bm B\cdot\hat{\bm n}$ errors on the plasma surface and at the exact same $15$m total coil length. 
\textcolor{black}{In Fig.~\ref{fig:force_scan} and all subsequent figures, we report all quantities in SI units, so that magnetic field quantities are reported in Tesla, forces are reported in Newtons, torques are reported in Newton-meters, and so forth.}

\begin{figure}
    \centering
    \includegraphics[width=0.99\linewidth]{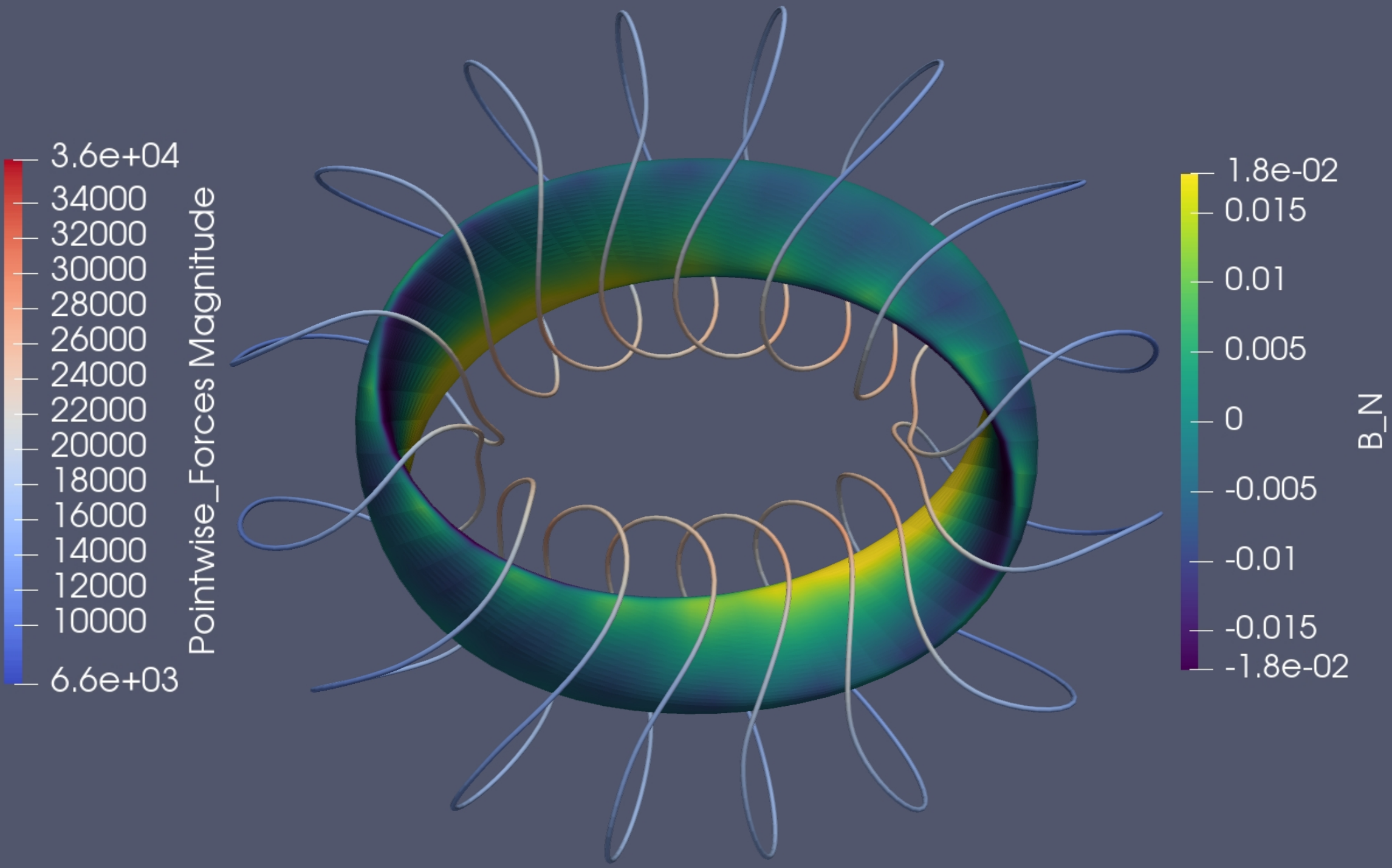}
    \caption{Minimizing pointwise forces with an appropriate threshold results in improved coil solutions.}
    \label{fig:force_thresholded}
\end{figure}

We find that pointwise and net forces cannot be reduced by more than a factor of two or so, at fixed coil length. As in Hurwitz et al.~\cite{hurwitz2024opt} force optimization tends to reduce the coil-plasma distance, especially in the inboard side, and with this extra ``slack'' in the coil lengths, a few coils reduced their forces by bending away from the plasma on the outboard side. These coils look geometrically complex, but this likely undesirable behavior can be mitigated by applying a threshold, e.g. $|d\bm F_i/dl_i|_0 = 10^4$ N/m. With this thresholded force term, the coils shown in Fig.~\ref{fig:force_thresholded} are substantially improved.

Starkly different results are shown for the net torques. The coil curvatures increase slightly and the changes are sufficient to decrease the net torque on every coil by several orders of magnitude, while the solution quality sees only minimal degradation. We also checked that the net-torque-optimized configuration has roughly equivalent forces on the coils as before, so that the torque reductions did not come at the cost of severe force increases. The fact that net torques can be easily reduced to negligible values without degradation to other objectives seems to hold up under large optimization scans described in the following text. These net torque reductions may be a useful fact for alleviating engineering complexity. The pointwise torques also see useful reductions when targeted; peak pointwise torques are reduced by about a factor of three after optimization. 

Although the primary objective in this manuscript is to use these objectives to show plausible dipole coil array solutions for three reactor-scale stellarators, there remain open questions about the robustness of the conclusions in this section about force and torque minimization across different stellarator devices, coil sets, and different initial conditions for optimization. Towards partially answering these questions, we performed several thousand ``cold start'' optimization runs with five coils per half-field period for each of the separate force and torque objectives, as in Hurwitz et al.~\cite{hurwitz2024opt}. Additional details about the generation of these illustrations can be found in that reference. 

\begin{figure*}
    \centering
    \includegraphics[width=0.9\linewidth]{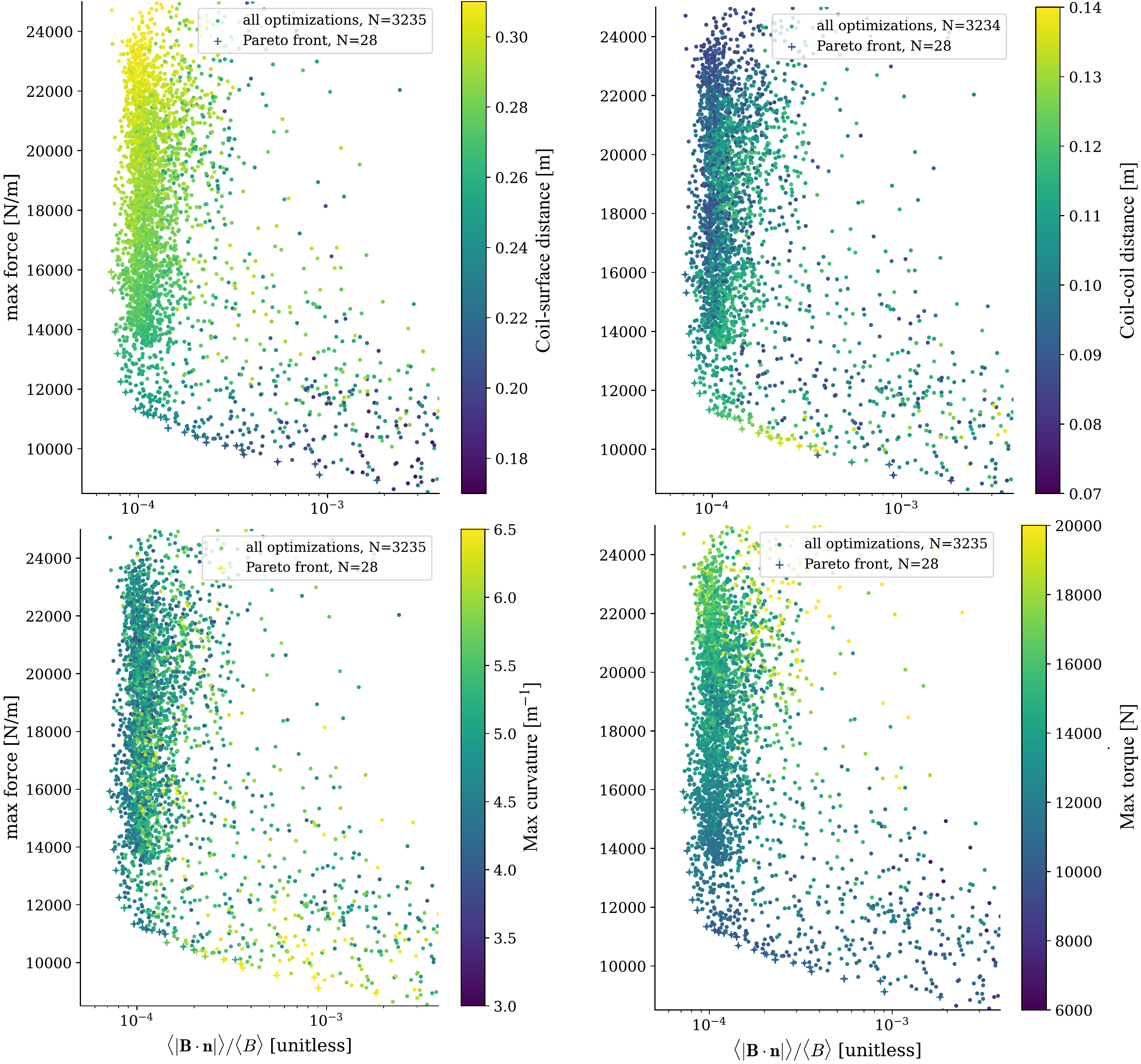}
    \caption{Summary of the strongest correlations found during Pareto scans of the pointwise forces. Top left: We reproduce the correlation seen in Hurwitz et al.~\cite{hurwitz2024opt} between larger coil-surface distances and larger maximum forces. Top right: the forces decrease with increased coil-coil distances, but low-force, high coil-coil distance solutions interestingly span the Pareto front. Bottom left: Larger maximum coil curvature seems to decrease at decreasing normalized field errors and increasing maximum coil force. Bottom right: Unsurprisingly, larger pointwise torques tend to correlate with larger maximum forces.}
    \label{fig:LPF_trends}
\end{figure*}

When optimizing only pointwise forces, we show in Fig.~\ref{fig:LPF_trends} that we can reproduce the correlation between larger maximum forces and larger coil-surface distances, as well as several other trends. In particular, maximum coil forces tend to decrease with larger coil-coil distances, decrease with higher coil curvature, and increase with higher maximum torques. Interestingly, there are a class of Pareto-optimal solutions that have low forces, low field errors, large coil-coil distances, but relatively larger curvature and smaller coil-surface distances. Each Pareto optimum describes a solution for which no other optimum is better in at least one regard~\cite{bindel2023understanding}. As far as we are aware, these are the best solutions found for the Landreman-Paul precise-QA. Despite using five TF coils per half-field period instead of four, they somewhat improve on the Wechsung et al.~\cite{wechsung2022precise} coils by reducing the total coil length and field errors while maintaining tolerable maximum forces.
Lastly, similar correlation  illustrations for pointwise torques, net forces, and net torques are illustrated in Appendix~\ref{sec:appendix_pareto} for completeness. 

\section{Results}\label{sec:results}
One of the primary benefits of the dipole coil arrays over permanent magnets is that they scale to high-field reactor-relevant scenarios. Therefore, we consider three stellarator configurations scaled to ARIES-CS parameters of average on-axis magnetic field strength $B_0 = 5.7$ Tesla and minor radius $r_0 = 1.7$m. 
Since in reactors a large neutron absorbent blanket will be required between the plasma and the coils, a minimum plasma-coil distance of $1.5$m is prescribed for all coils. The minimum TF coil-coil distance is given as $0.8$m and we allow a minimum distance of half this value between the TF coils and the dipole coils. 

This setup assumes that the dipole array can be installed and designed such that it can be easily removed during maintenance, but we leave these engineering considerations to future work and proceed primarily with the question of feasibility. Typical estimates of maximum force loads for VIPER high-temperature superconducting (HTS) cables~\cite{hartwig2020viper} are $\sim 400$ kN/m~\cite{zhao2022structural,riva2023development}. SPARC toroidal field coils had maximum force loads of $\sim 800$ kN/m~\cite{hartwig2023sparc} although they use noninductive coils. Since the maximum force loads are approximate, we will simply search for solutions with maximum force loads $\lesssim 1$ MN/m. The radii of the reactor-scale design dipole coils in this work are $\sim 1$ m, so that, distributed throughout the circumference, the maximum tolerable net force on the dipoles is estimated as $\sim 6$ MN. Similarly, the TF coil(s) will have length(s) $\sim 40-60$ m, with a maximum tolerable net force $\sim 50$ MN. Adopting the SPARC number, we assume $200$ turns for the TF coils and $100$ turns for the dipole coils. If this fixed number of turns is sufficient to keep forces in material limits, it can be straightforwardly reduced to alleviate the high cost of purchasing long lengths of superconductor.\footnote{Throughout this work, the amount of superconductor or HTS tape required is assumed to be dictated by the forces since we are not modeling the critical current. In general, the maximum dipole currents dictate the number of winds required for a given critical current. At a certain point, the finite build of each coil may make it challenging to wind small dipole coils.
\textcolor{black}{The planar coils in this work all have roughly 1 m major radius, and a 5$\times$5 cm rectangular cross section is assumed to compute the self-forces. We assumed for simplicity that 100 turns are used for the planar coils, so that 10 MA / 100 turns gives 100 kA in a single turn of wire. We now use some estimates from Molodyk et al.~\cite{molodyk2021development}. 
The critical current densities for this class of rectangular cross section superconducting wires at $B \sim 6$ T and $ T \sim 4.2$ K is of order 1 kA/mm$^2$. Using enough wire for a cross-section of 1 cm $\times$ 1 cm is then sufficient to achieve 100 kA per “winding pack” with 100 turns, for a total volume of 100 cm$^2$. So there is a slight mismatch here, but the optimized dipole coils are kept far enough away from each other to increase the cross-sectional area to a 10$\times$10 cm rectangular cross-section. 
This is consistent with assumptions made in Kruger et al.~\cite{kruger2025coil}, where they cap the currents at 1.6 MA, with half the estimate above (0.5 kA/mm$^2$) for the critical current density. 
}
}
Estimates are not obviously available for net torques, but using the major radius $R_0 \sim 1$m of a dipole coil as a characteristic scale, we consider the maximum net torque to be $\sim 6$ MN-m. Similarly, for the TF coils, we consider a maximum net torque of $\sim 400$ MN-m.

The locations of the dipole coils are initialized by taking an inner and outer toroidal surface, both determined by extending the plasma surface using the normal vector to the plasma surface times a fixed offset distance. Then coils are uniformly initialized in between this inner and outer toroidal surface with spacing between the coils large enough that coil-coil intersections are avoided in the initial configuration. The coil array is initialized at $1.5$m or further away from the plasma surface. Coils interlinking with the initial TF coils are simply removed before optimization starts. For geometric simplicity and potential mass production, coil shapes are fixed to be planar circular coils, although this is not a requirement of the optimization problem. 

If the details are not of particular interest, we provide a summary of the final three reactor-scale dipole array solutions produced in this work  in Table~\ref{tab:dipole_arrays_summary}.

\begin{table*}
    \centering
    \begin{tabular}{|c|c|c|c|c|} \hline 
         &Landreman-Paul QA & Landreman-Paul QH &  Schuett-Henneberg QA \\ \hline 
         \# of unique TF coils &  3 & 2 & 2 \\ 
\hline
\textcolor{black}{
         Aspect ratio} & \textcolor{black}{6.00} & \textcolor{black}{8.00} & \textcolor{black}{2.42} \\ 
\hline
         \# of unique dipole coils  & 41 & 27 & 16 \\ \hline
         Total TF coil length &  $115.2$ m & $90.1$ m &  $75.5$ m \\ 
         \hline
         Maximum dipole current &  $14$ MA & $9.5$ MA &  $16$ MA \\ \hline
         Maximum dipole force &  $1.0$ MN/m & $0.67$ MN/m &  $1.3$ MN/m \\ \hline
         Maximum TF force &  $0.51$ MN/m & $0.81$ MN/m &  $0.61$ MN/m \\ \hline
         Maximum dipole net torque &  $5.0$ MN-m & $5.7$ MN-m &  $7.1$ MN-m \\ \hline
         Maximum TF net torque &  $55$ MN-m & $130$ MN-m &  $38$ MN-m \\ \hline
         $\langle \bm B \cdot\hat{\bm n} \rangle / \langle B\rangle$ &  $9.0 \times 10^{-4}$ & $6.0
        \times 10^{-4}$ &  $1.9\times 10^{-3}$ \\ \hline
         $f_\text{QS}$ &  $2\times 10^{-4}$ & $5\times 10^{-4}$ &  $1.3 \times 10^{-3}$ \\ \hline
    \end{tabular}
    \caption{Summary of the finalized dipole array designs for the three stellarators considered in the present work. \textcolor{black}{The stellarator configurations are scaled to ARIES-CS parameters of average on-axis magnetic field strength $B_0 = 5.7$ Tesla and minor radius $r_0 = 1.7$m.} }\label{tab:dipole_arrays_summary}
\end{table*}

\subsection{Reactor-scale Landreman-Paul QA stellarator}\label{sec:results_QA}
Quasi-axisymmetric (QA), low-field period stellarators are a top choice for future stellarator reactors because of their relative geometric simplicity, but QA stellarators can exhibit large bootstrap currents that complicate stellarator optimization and experimental operation. The Landreman-Paul QA configuration is stellarator symmetric, two-field period symmetric, and exhibits a very high-degree of quasi-axisymmetry. 

For the following dipole array solution, we consider three unique TF coils. Simple and smooth TF coils are prescribed by limiting the order of the coils to $M' = 4$ and penalizing the length as much as is consistent with the other objectives.

\subsubsection{\textcolor{black}{Reactor-scale dipole array solution}}\label{sec:reactorscale_QA}
We begin by illustrating force and torque optimized results using a dipole array. We achieve essentially the same normalized error $\langle \bm B \cdot\hat{\bm n}\rangle / \langle B\rangle \approx 3.4\times 10^{-3}$ than the (rescaled) four coil solution in Wechsung et al. with $182.3$m total coil length, but use three shorter TF coils with lengths 33.6, 33.8, and 35.6 meters for a total of 103.0 m. There are 41 unique dipole coils, 164 dipole coils in total, initialized to random orientations, each with major radius $0.743$m. All of the coil constraints are satisfied in the final solution; none of the coils are interlinking, all TF-TF distances are 0.8 m or larger, all TF-dipole distances are $0.44$m or larger, and all coils are at least $1.5$ m from the plasma surface. The curvature of the TF coils is very mild because only four Fourier modes are used for the parametrization of the TF curve. We verify that the accuracy is sufficient by producing Poincaré plots and constructing a quadratic flux-minimizing (QFM) surface~\cite{dewar2010unified} with the obtained coil solution. A volume constraint is used, and the QFM is considered to be converged if the normalized quadratic flux $\leq 10^{-6}$.
This QFM is fed into the 3D MHD equilibria code VMEC~\cite{hirshman1983steepest} and the two-term quasisymmetry error $f_\text{QS}$ is computed via,
\medmuskip=-1mu
\thinmuskip=-1mu
\thickmuskip=0mu
\begin{align}
    f_\text{QS} = \sum_s\left\langle \left(\frac{(N - \iota M)\bm B \times \nabla B\cdot\nabla\psi - (MG + NI)\bm B \cdot\nabla B}{B^3}\right)^2\right\rangle.
\end{align}
\medmuskip=4mu
\thinmuskip=4mu
\thickmuskip=4mu
Above, $\iota$ is the rotational transform, $G(s)$ is proportional to the poloidal current outside the flux surface labeled by the normalized toroidal flux coordinate evaluated at $s = 0, 0.1, ..., 1$, $I(s)$ is proportional to the toroidal current inside the surface, $\psi$ is proportional to the toroidal flux, $\langle .\rangle$ brackets denote average over a flux surface, and $(M, N)$ are integers that specify the  helicity of symmetry. Note that $f_\text{QS}$ has no units. 

We find that our coil solution has errors $f_\text{QS}\approx 5.0\times 10^{-3}$. This is increased substantially from $f_\text{QS}\approx 1.6\times 10^{-6}$ in the original configuration, but still represents reasonably good quasisymmetry that is comparable to or better than the present stellarators in operation. 
The reduced value may be in part explained by the point raised in Rodriguez et al.~\cite{rodriguez2022measures} and Wiedman et al.~\cite{wiedman2023coil} that deviations from quasisymmetry with higher mode numbers, such as those created by modular coil ripple or small dipole coils, are weighted more strongly in the two-term measure.

\begin{figure}
     \centering
     \includegraphics[width=\linewidth]{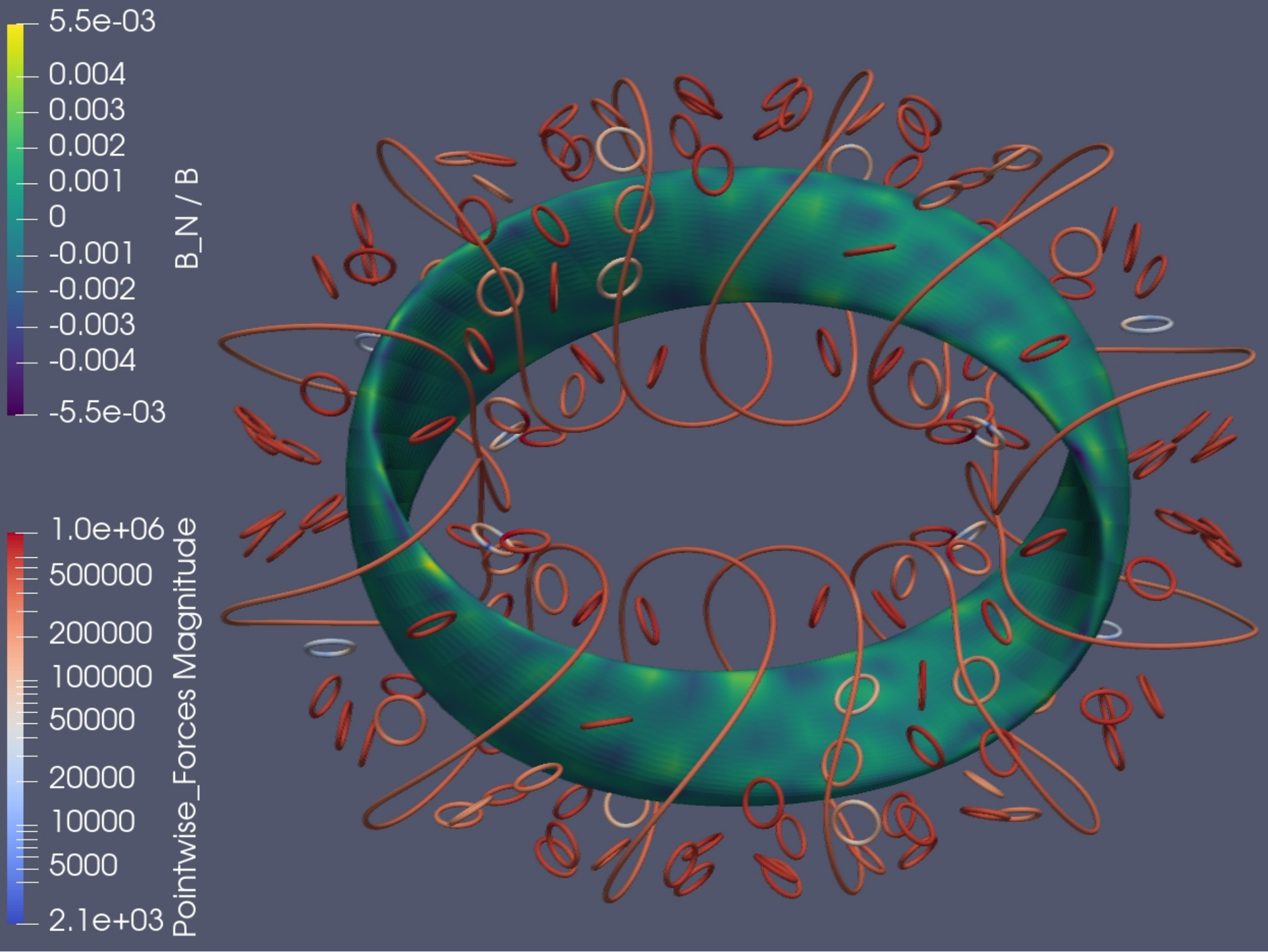}
     \caption{Dipole array design for the QA stellarator with precise quasisymmetry.}
     \label{fig:QA_refined}
 \end{figure}

In order to attain a coil solution with much improved quasisymmetry levels, we initialize a second optimization using the force and torque optimized dipole array coil configuration from above as an initial condition. We relaxed the penalty slightly on the TF coil lengths in order to achieve a reduced $\langle\bm B \cdot\hat{\bm n}\rangle / \langle B \rangle
 \approx 9.0\times 10^{-4}$ with TF coils of total length $115.2$m ($110$m of total TF coil length is sufficient but another $5$m further minimizes the field errors and the net torques). Using a QFM, we calculate the quasisymmetry error is now $f_\text{QS} \approx 1.5\times 10^{-4}$, representing very precise levels of quasisymmetry.  Figure~\ref{fig:QA_refined} illustrates the pointwise forces and field errors on the plasma surface. The maximum forces are $1.0$ MN/m and the maximum net torques are $5.0$ MN-m, both roughly within the material tolerances. 
This solution represents the first dipole array solution with tolerable forces, torques, coil-coil distances, etc. for a reactor-scale stellarator.

 \begin{turnpage}
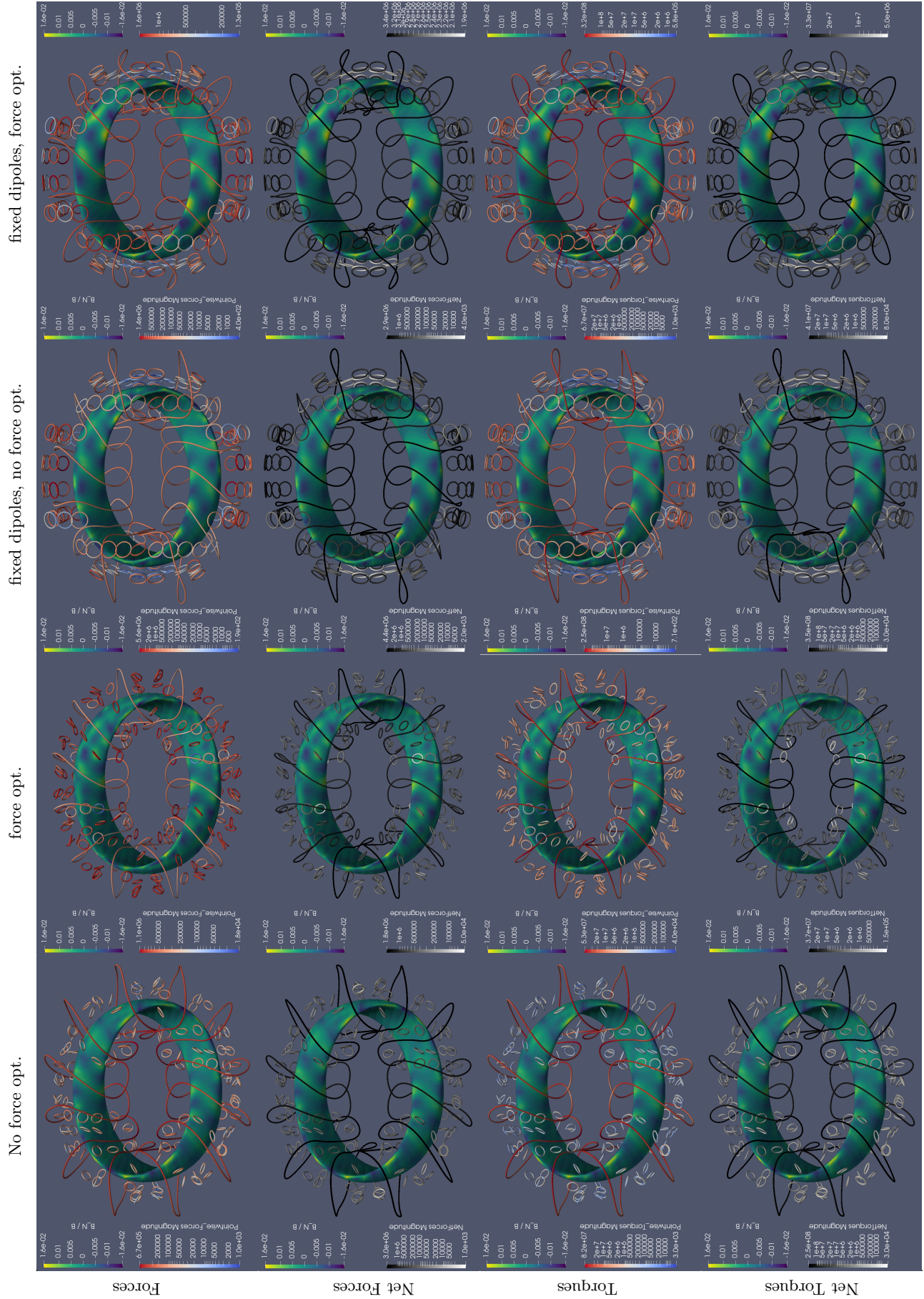
\begin{figure*}
    \centering
\begin{overpic}[width=\linewidth]{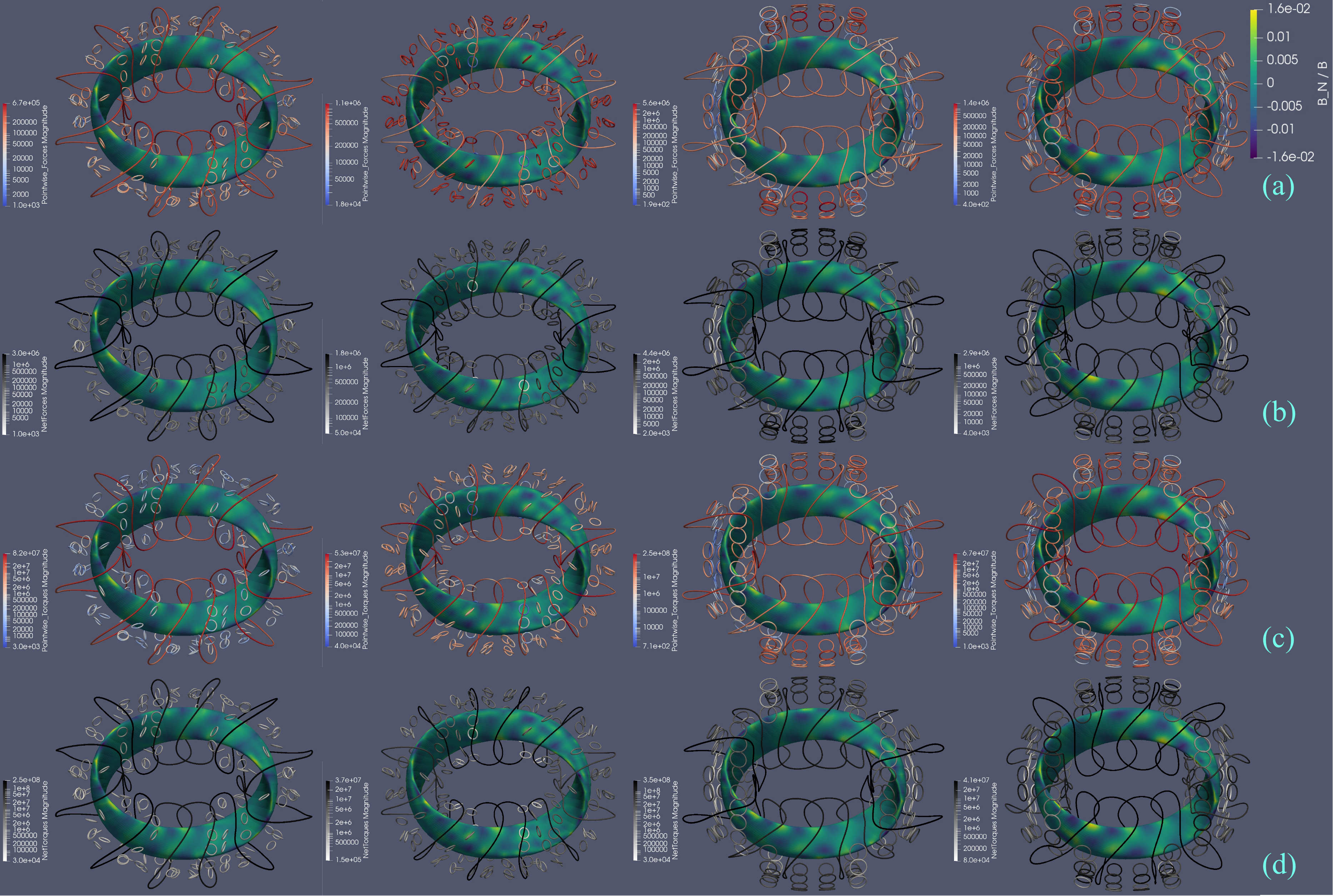}
\put(9, 68){No force opt.}
\put(34, 68){force opt.}
\put(54, 68){fixed dipoles, no force opt.}
\put(81, 68){fixed dipoles, force opt.}
\put(-1, 57){\begin{rotate}{90}Forces\end{rotate}}
\put(-1, 39){\begin{rotate}{90}Net Forces\end{rotate}}
\put(-1, 23){\begin{rotate}{90}Torques\end{rotate}}
\put(-1, 3){\begin{rotate}{90}Net Torques\end{rotate}}
    \end{overpic}
    \caption{Left to right: Solutions (1) with dipole coils without force/torque optimization, (2) dipoles with force/torque optimization, (3) fixed dipoles without force/torque optimization and (4) fixed dipoles with force/torque optimization. Total TF coil lengths are $123.4$m, $103.0$m, $119.8$m, and $123.3$m, respectively. Top to bottom visualizes (a) pointwise forces, (b) net forces, (c) pointwise torques, and (d) net torques.}
    \label{fig:QA_baseline}
\end{figure*}
\end{turnpage}

\subsubsection{\textcolor{black}{Ablation study}}\label{sec:ablation_study}
After obtaining a suitable reactor-scale dipole array configuration with tolerable forces and torques, we pursued an ablation study where we sequentially remove coils, terms in the objective, or degree of freedoms in the coils, and see how the solution changes. 
We performed the following optimizations: we (1) optimize the dipole coils, (2) optimize the dipole coils with force and torque objectives included $-$ this is the solution already presented in the previous Sec.~\ref{sec:reactorscale_QA}, (3) fix the coil locations and orientations in space, (4) repeat the fixed locations and orientations, but remove all the force and torque objectives, and finally (5) remove all the dipole coils. 

For (3) and (4), we initialize the orientations to point towards the nearest point on the plasma surface. We also found empirically that minimizing the forces and torques on dipole coils in the tight inboard access areas was very challenging, so we changed the grid. Now $36$ dipole coils with radii $1.09$m are initialized further out at $2.25$m away from the plasma, and some of the inboard coils are removed before optimization. For (5), we needed to increase the number of Fourier modes from four to twenty in order to achieve comparable field errors. We also attempted to find dipole array solutions with a single unique TF coil but found that the single coil still needed to be quite long, and the peak forces were difficult to minimize within material tolerances. All of the optimizations (1)-(5) are performed to produce a similar $\langle \bm B \cdot\hat{\bm n}\rangle / \langle B\rangle\sim 3\times 10^{-3}$, relaxing the constraint on the total coil length if needed. At these error levels, Poincaré plots seem to reliably reproduce the expected nested flux surfaces. 

\begin{figure}
    \centering
    \includegraphics[width=0.72\linewidth]{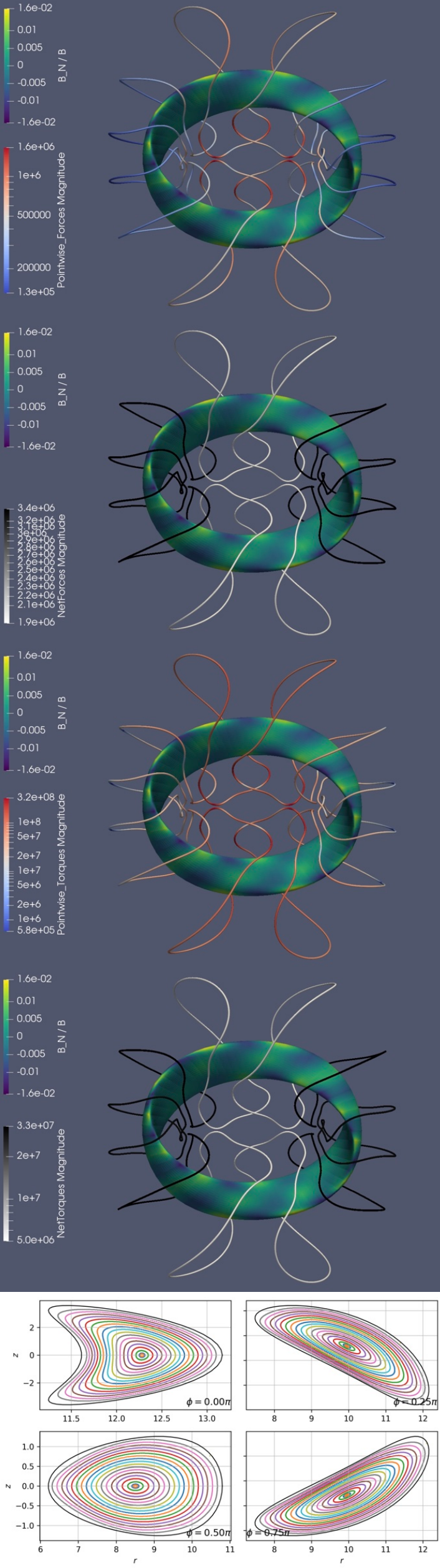}
    \caption{Similarly accurate coil solution with two TF coils and no dipole coils requires complex coils with $146.8$m total length. \textcolor{black}{The outermost black line in the Poincaré plots indicates the target plasma boundary.}}
    \label{fig:QA_TFonly}
\end{figure}

A detailed comparison of optimization results (1)-(4) is shown in Fig.~\ref{fig:QA_baseline} and the TF-only solution is shown in Fig.~\ref{fig:QA_TFonly}. 
The total TF coil lengths for each of (1)-(5) are, respectively,  $123.4$m, $103.0$m, $119.8$m, $123.3$, and $146.8$m. The maximum dipole currents for each of (1)-(4) are, respectively,  $4.1$ MA, $15$ MA, $35$ MA, and $17$ MA.
The takeaways are:
\begin{itemize}
    \item With fixed dipole coil locations and orientations, solutions are found with very large forces and torques, unless these terms are directly minimized. In this case, solutions with tolerable forces and torques can be found. 
    \item In all cases with dipole arrays, the net torques need to be directly minimized, and this can be effectively done. 
    \item By minimizing pointwise forces and net torques directly, the total TF coil lengths from optimization (1) to (2) are reduced by more than 20 meters at the same time that peak dipole currents increase by an order of magnitude, net torques reduce by an order of magnitude, and maximum pointwise forces stay within a factor of two. In other words, directly optimizing forces and torques facilitates substantial reductions in the TF coil lengths and dipole coils that contribute much stronger magnetic fields to the solution.
    \item Despite TF coil lengths of just $103$m in optimization (2), the total amount of high-temperature superconducting tape for the TF and dipole coils increase by 10 km compared to the no dipole solution with TF coils of $146$m (40 km after applying discrete symmetries, a $35\%$ increase). However, it only slightly increases the amount of superconductor compared with the scaled-up solution from Wechsung et al.~\cite{wechsung2022precise}. Moreover, the amount of superconductor can likely be reduced as we see plenty of variation on this point, and in the next Sec.~\ref{sec:results_QH} we find that the dipole solution uses drastically \textit{less} superconductor than the corresponding published reactor-scale coil set.
    \item In the fixed dipole cases (3) and (4), there are some dipole coils in the array that experience very low forces and torques because they contribute very little to the magnetic field and are fixed in a relatively unimportant location to reduce $\bm B\cdot\hat{\bm n}$ errors. These coils could be removed with sequential thresholding algorithms~\cite{kaptanoglu2023sparse}, or at least the number of wire turns could be drastically reduced. These order-of-magnitude differences between forces and magnetic field contribution between dipole coils shrink significantly when the dipole coils can move and orient in space.
\end{itemize}

For configurations with fixed dipole coils, coils that conform to a toroidal surface without gaps or overlaps, such as in a winding surface method~\cite{ku2010modular} or the wireframe technique from Hammond~\cite{hammond2024framework}, may be more advantageous. However, these designs are beyond the scope of this work, as they have so far required nonplanar coils, a potential disadvantage when it comes to mass production.

We conclude by noting that this is not definitive proof that better coil configurations, with or without dipole arrays, cannot be found with further rounds of optimization. Adding a fourth unique TF coil could also help, although there are diminishing returns, since constraints such as the minimum coil-coil distance can become quite restrictive on the inboard side of the plasma surface. 

We conclude this ablation study by noting the main takeaways: reactor-scale devices can be built with dipole arrays exhibiting tolerable forces and torques, net torques can be dramatically reduced during optimization with minimal tradeoffs, and ablation studies indicate that providing geometric degrees of freedom to the dipole arrays may sometimes be a pre-requisite for finding plausible or improved reactor-scale solutions.

\begin{figure*}
    \centering
    \begin{overpic}[width=\linewidth]{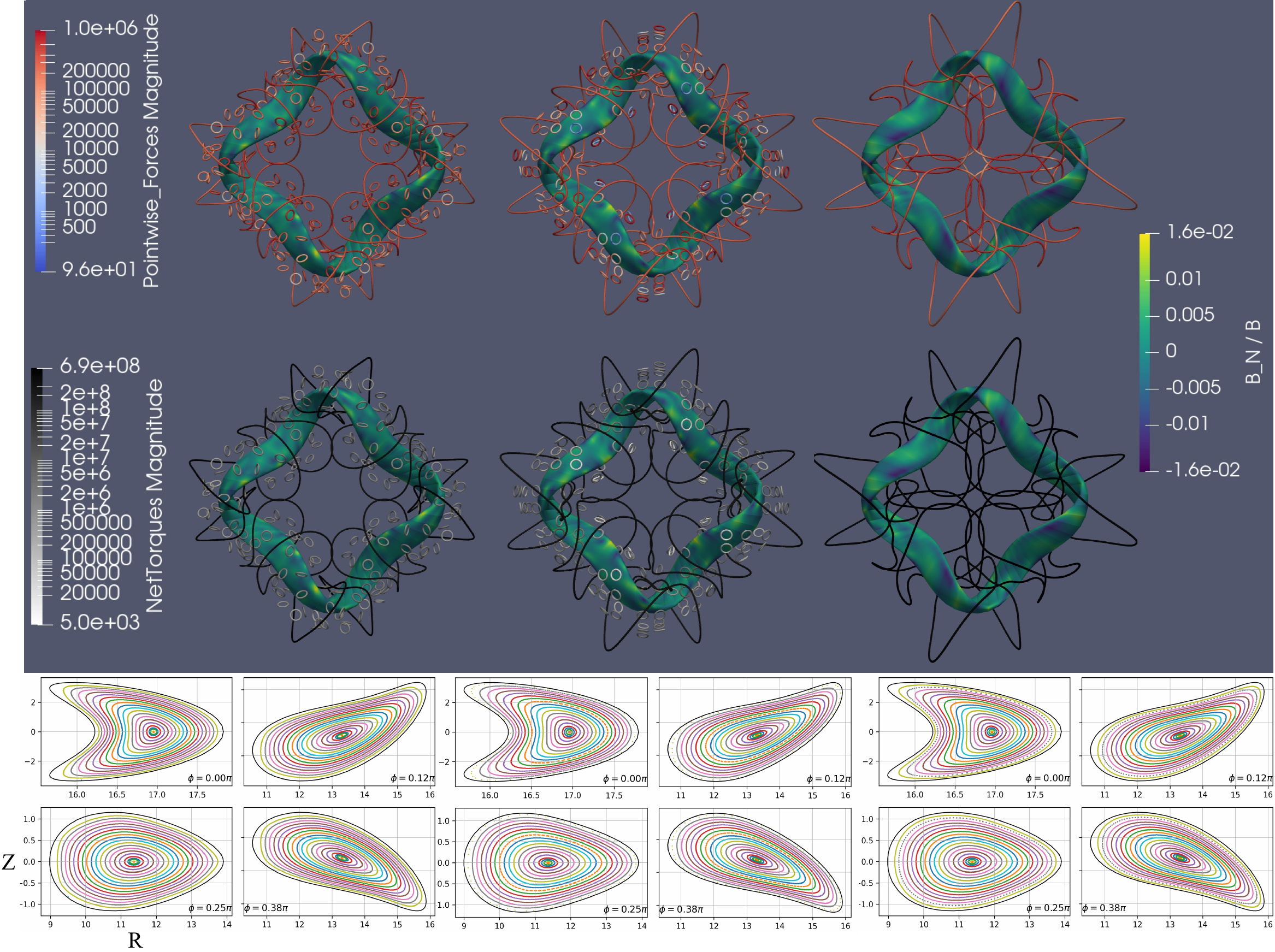}
    \put(20, 76){Dipole coils}
    \put(43, 76){Fixed dipole coils}
    \put(70, 76){No dipole coils}
    \put(1, 63){\begin{rotate}{90}Forces\end{rotate}}
    \put(1, 33){\begin{rotate}{90}Net Torques\end{rotate}}
    \put(1, 10){\begin{rotate}{90}Poincaré\end{rotate}}
    \end{overpic}
    \caption{Summary of the dipole coil optimization results with force and torque optimization for the Landreman-Paul QH stellarator, showing field errors, pointwise forces, net torques, and Poincaré plots. Peak forces per unit length and peak net torques are: left column $8.5\times 10^5$ N/m and $4.2\times 10^8$ N-m, center column $1.0\times 10^6$ N/m and $2.0\times 10^8$ N-m, right column $7.9\times 10^5$ N/m and $6.9\times 10^8$ N-m. \textcolor{black}{The outermost black line in the Poincaré plots indicates the target plasma boundary.}}
    \label{fig:QH_solution}
\end{figure*}

\begin{figure}
    \centering
    \includegraphics[width=0.915\linewidth]{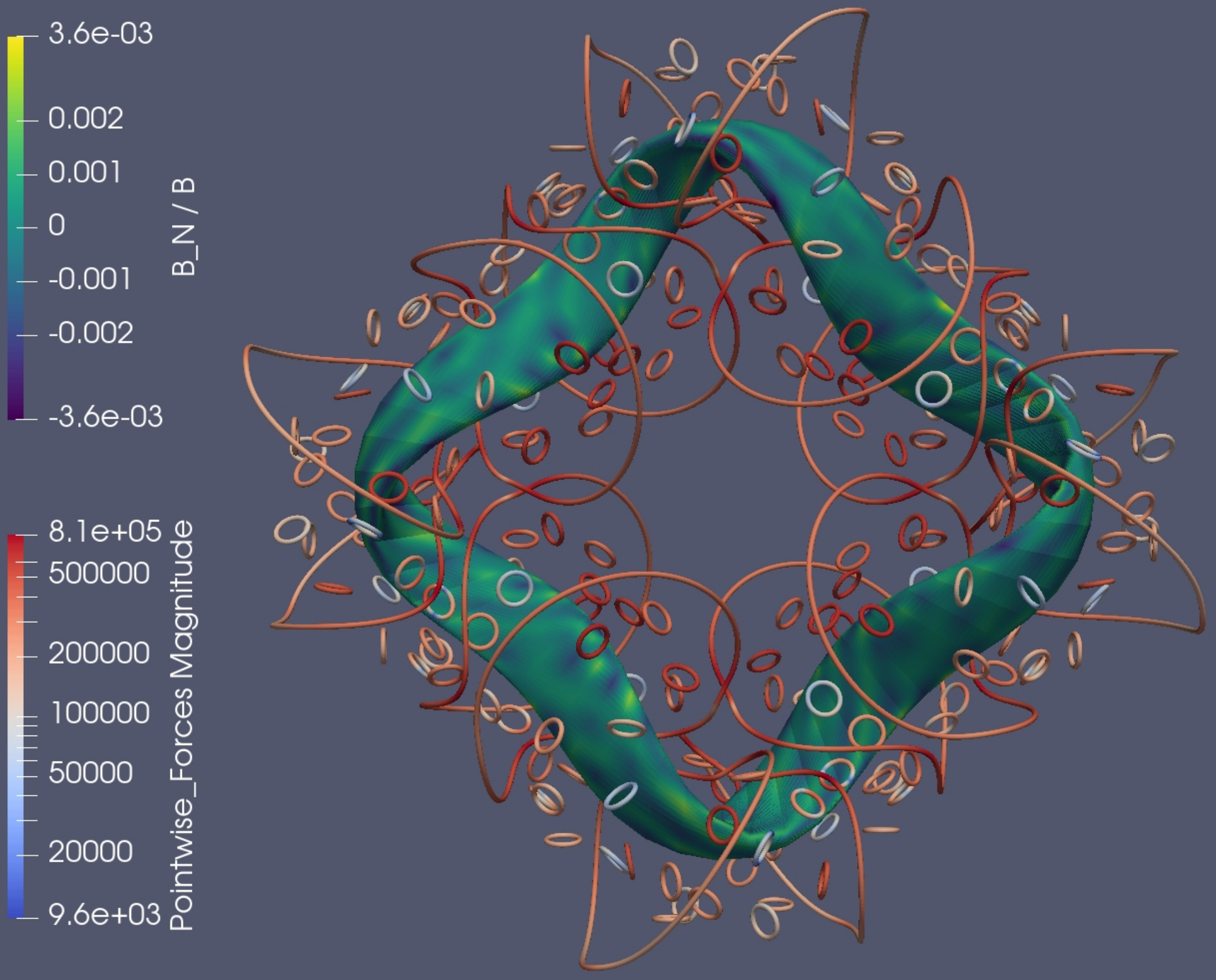}
    \caption{QH coils obtained after a second  optimization with total TF coil length increased to $90.1$m.}
    \label{fig:QH_2ndopt}
\end{figure}

\subsection{Reactor-scale Landreman-Paul QH stellarator}\label{sec:results_QH}
It is likely that the plausibility of dipole coil array solutions is strongly dependent on the shape of the plasma surface. 
For this reason, we now illustrate that many of the conclusions presented in this work are not specific to the Landreman-Paul QA stellarator.
Quasi-helical (QH) stellarators have a number of important differences from other stellarator classes, such as reduced bootstrap current compared to quasiaxisymmetric stellarators. Unfortunately, QH stellarators often need quite complex modular or helical coils compared to QA stellarators. The only optimized quasi-helical stellarator that has been built is HSX and it uses \textit{six} modular coils per half-period~\cite{anderson1995helically,almagri1999helically}. 

Recently, QH stellarators have been found numerically that exhibit a very high degree of quasisymmetry.
We will consider this vacuum precise-QH solution obtained by Landreman and Paul~\cite{landreman2022magnetic}. The configuration is stellarator-symmetric and four-field-period symmetric. Modular coil solutions for this reactor-scale stellarator have been found in Wiedman et al.~\cite{wiedman2023coil}, helical solutions have been found in Kaptanoglu et al.~\cite{kaptanoglu2024topology}, and permanent magnet solutions (for a table-top scale experiment) have been found in Kaptanoglu et al.~\cite{kaptanoglu2023sparse}. 
The modular coils in Wiedman et al.~\cite{wiedman2023coil} consisted of 5 coils per half-period, each represented by 7 Fourier modes, with average coil length $35.56$m, total coil length of $177.76$m, minimum coil-coil distance $1.09$m, minimum coil-surface distance of $1.62$m, maximum curvatures between 0.6 m$^{-1}$ and 0.8 m$^{-1}$, and average mean square curvature $0.069$ m$^{-1}$. An average normalized $\langle\bm B\cdot\hat{\bm n}\rangle/\langle B\rangle \approx 6\times 10^{-4}$ was achieved. 

In contrast, our final coil design uses 27 dipole coils per half-period, each with radius $0.792$m, and only 2 TF coils per half-period, each parameterized by only four Fourier modes. As mentioned earlier, the number of Fourier modes describing the TF coils is so restrictive that penalties on the coil curvature, torsion, etc. are typically unnecessary. The two coils are $40.2$m and $42.0$m long respectively with total length $82.2$m, and satisfy all the coil constraints: final minimum TF-TF coil distance $0.79$m, final minimum TF-dipole distance of $0.56$m, minimum coil-surface distance of $1.5$m, and so on. In other words, we use 24 fewer TF coils than the Wiedman design, and our coils have coil lengths, coil-coil distances, and coil-plasma distances similar to those of the Wiedman design. The average normalized field errors are four times higher, at $\langle \bm B \cdot\hat{\bm n}\rangle /\langle B \rangle \approx 2.8\times 10^{-3}$, but Poincaré plots in Fig.~\ref{fig:QH_solution} illustrate that the flux surfaces are very well-preserved.
Assuming that the Wiedman et al. coils also require $\sim 200$ turns to reduce the forces or stay within the critical current limit, and after accounting for all the coils via the discrete symmetries, the dipole solution reduces the amount of high-temperature superconducting tape by \textcolor{black}{about 16\%}, or about 45 km!

For a point of comparison, we attempted two additional optimizations: (1) to find an optimal two TF-coil solution without dipole coils that achieves comparable errors and (2) to fix all the dipole locations and orientations. The three optimizations are shown in Fig.~\ref{fig:QH_solution}.

In (1), we found that achieving comparable errors is possible only by increasing the total coil lengths to $\sim 108$m.  However, then the forces and torques on the TF coils exceed the tolerances by orders of magnitude. Repeating the same experiment until the forces and torques were roughly within tolerances, and the coil curvatures were within a factor of two of those in Wiedman et al., we obtain a solution with a total coil length of $120$m. We also tried increasing the number of Fourier modes from four to twenty, but could not find any solutions that were not very geometrically complex; we settled on using eight Fourier modes. The TF-coil-only solution actually uses slightly less total superconducting material than the dipole solution, although visualization in Fig.~\ref{fig:QH_solution} shows that the geometric complexity of such coils remains very high. The takeaway here is really that an equivalent solution with two plausible TF coils, and no dipole coils, seems unlikely to exist.  

For (2) we use a slightly modified dipole array grid of 25 dipoles per half field period with radii $0.80$m. We found that coils must be initialized using a toroidal surface $1.75$m away from the plasma surface to guarantee that the minimum coil-plasma distance $d_{cs} \geq 1.5$m because dipoles are initialized to orient towards the nearest plasma point. As in Sec.~\ref{sec:ablation_study}, we find that minimizing the pointwise forces and net torques is imperative to obtaining a plausible solution. Moreover, the total TF coil length increases substantially to $101.4$m, now saving $21.6$km of tape compared to the Wiedman solution.

Finally, while the Poincaré plots show the magnetic surfaces are quite well-preserved, we repeat the analysis in the previous section of constructing a QFM surface from the obtained coil solution and recomputing the average two-term quasisymmetry $f_\text{QS}$.
We find that the quasisymmetry error is increased from $f_\text{QS} = 4.0\times 10^{-5}$ to $f_\text{QS} = 2.4 \times 10^{-2}$. This is at a value where the fast-ion losses become non-negligible and we performed an additional optimization to mitigate this. We use the force and torque optimized dipole array solution as an initial condition for an additional optimization with TF coils slightly increased to $90.1$m, still saving $32$km of superconductor compared to the published modular coil solution. We achieve an improved normalized error $\langle \bm B \cdot\hat{\bm n}\rangle / \langle B\rangle=7.8\times 10^{-4}$, average coil curvature $\kappa \approx 0.76$m$^{-1}$, average mean-squared curvature $0.06$m$^{-2}$, and construct a QFM to compute an average two-term quasisymmetry error of $5 \times 10^{-4}$. The solution illustrated in Fig.~\ref{fig:QH_2ndopt} now has very good quasisymmetry, near the value obtained in the Wiedman et al.~\cite{wiedman2023coil} solution with 5 coils per half-period that found essentially zero fast-ion losses. Notably, we were unable to find a similarly performant solution when the dipole coils were fixed in space and orientation, even when the TF coil lengths were significantly increased and coil forces made substantially larger.

\begin{figure*}
    \centering
    \includegraphics[width=\linewidth]{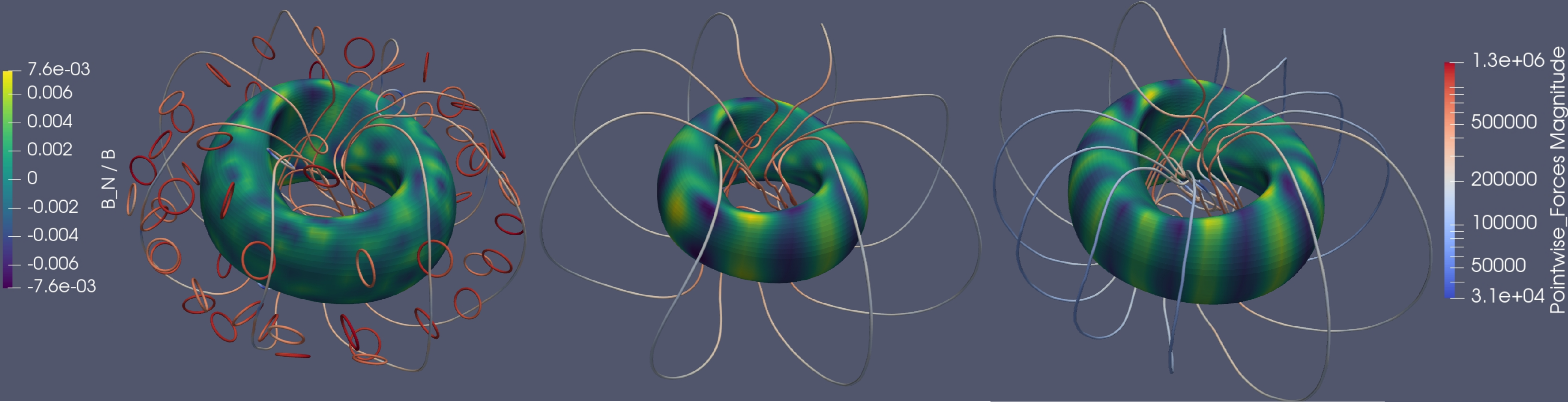}
    \caption{Left: Dipole coil solution for the compact Schuett-Henneberg QA stellarator with 16 dipole coils per half field period, total TF coil length $75.5$m; Center: two TF coil solution total TF coil length $95.5$m; Right: three TF coil solution, total TF coil length $110.6$m. Notice the dipole coils mitigate the modular coil ripple.}
    \label{fig:QA_henneberg}
\end{figure*}

\begin{figure}
    \centering
    \includegraphics[width=\linewidth]{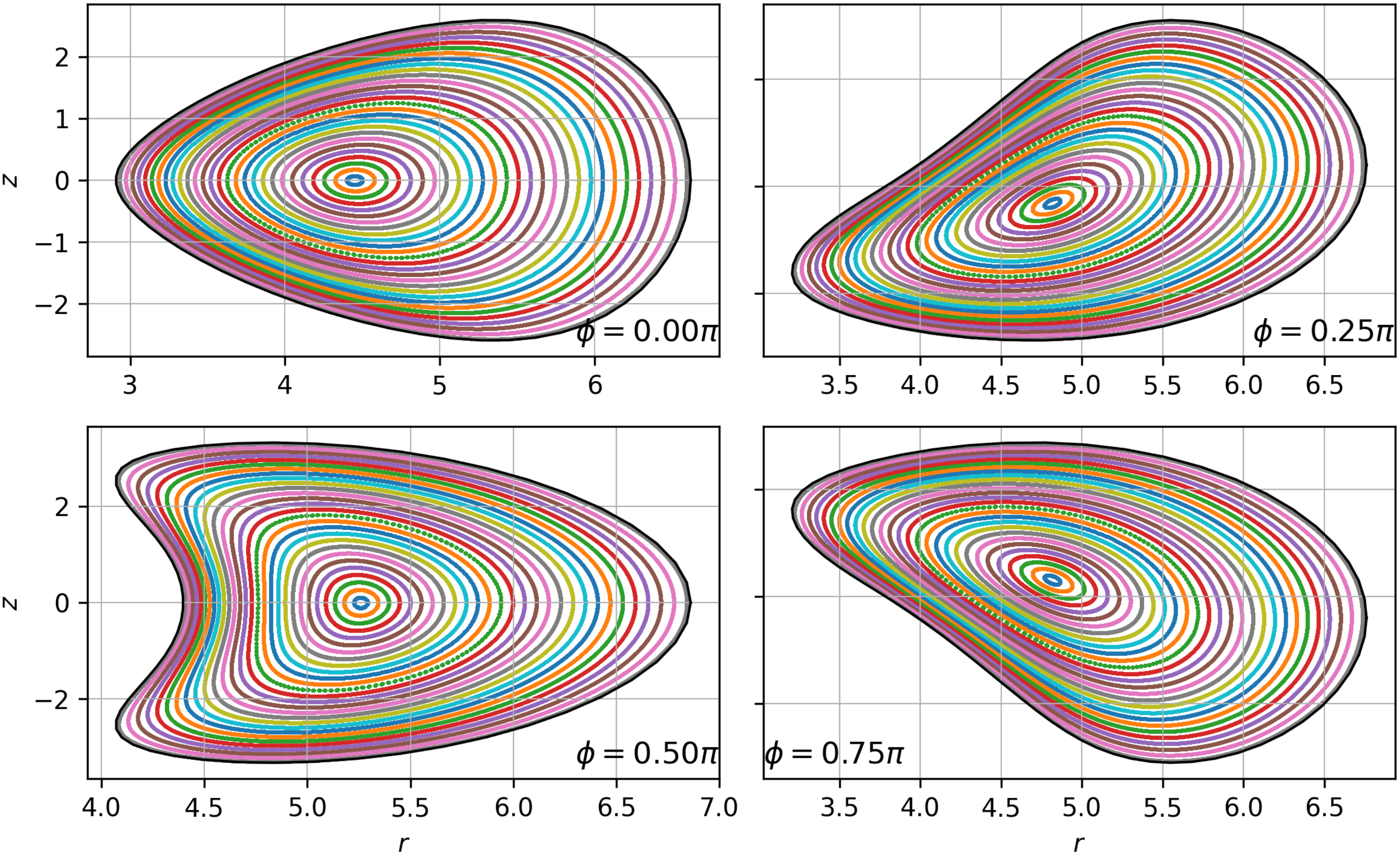}
    \caption{Poincaré plots for the Schuett-Henneberg two-field period compact QA stellarator verify the accuracy of the dipole array. \textcolor{black}{The outermost black line indicates the target plasma boundary.}}
    \label{fig:henneberg_poincare}
\end{figure}
\begin{figure}
    \centering
    \includegraphics[width=\linewidth]{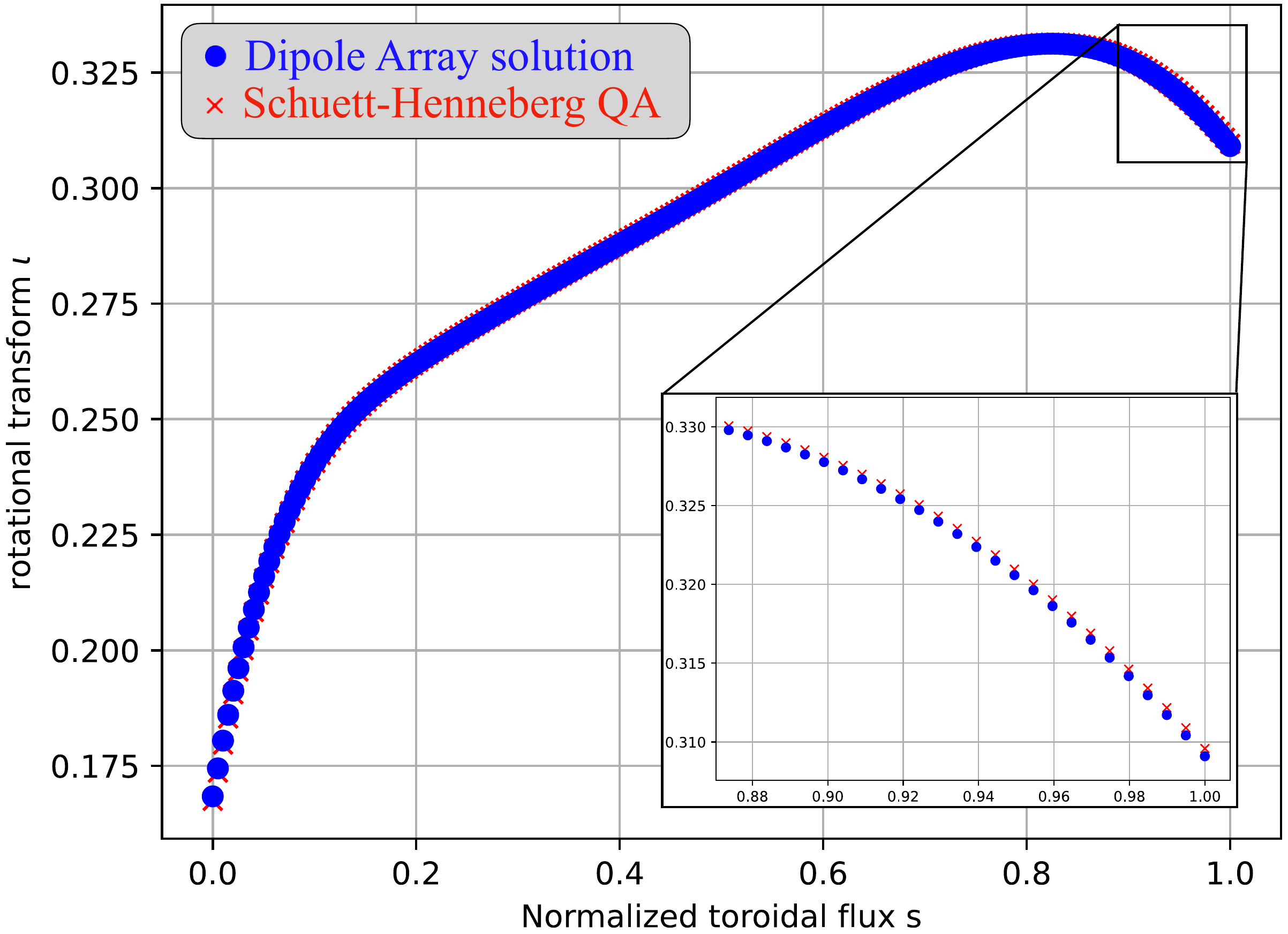}
    \caption{\textcolor{black}{Rotational transform profile for the Schuett-Henneberg dipole array solution is essentially indistinguishable from the original design without coils. }}
    \label{fig:henneberg_iota}
\end{figure}

\subsection{Reactor-scale Schuett-Henneberg QA stellarator}
Lastly, we find a dipole coil solution for the two-field-period Schuett-Henneberg QA stellarator~\cite{schuett2024exploring}. This is a substantially more compact device than the Landreman-Paul QA stellarator; the major radius is decreased by more than a factor of two from $10.13$m to $4.89$m. Subsequently, it is difficult to place any dipole coils on the inboard side of the plasma. Instead, we use a dipole array to mitigate the outboard modular coil ripple. This allows us to reduce the number of required TF coils, which in turn increases the amount of space available in the inboard side. We find a reactor-scale solution with only two TF coils with total length $75.5$m. We let the TF coil representation include sixteen Fourier modes to minimize inboard field errors and add back the objective terms related to coil curvature, requiring curvature below $1.0$m$^{-1}$ and mean-squared curvature $\leq 0.15$m$^{-2}$, comparable or better than the stellarator coil solutions listed in Table 1 of Wiedman et al.~\cite{wiedman2023coil}.

The dipole array is initialized at $1.5$m from the plasma surface, as described above, and then we remove any coils on the inboard side or interlinking with the initial TF coils. In addition, we leave a gap in the dipole coils at $\zeta = \pm \pi/2$ and remove most coils above and below the plasma for additional access. There are $16$ dipole coils per half-field period, each with radius $0.783$m. The solution with $\langle \bm B \cdot\hat{\bm n}\rangle / \langle B \rangle \approx 1.9\times 10^{-3}$, TF coil average curvature $\kappa\approx 1$m$^{-1}$, and TF coil average mean square curvature $0.15$m$^{-2}$ is illustrated in Fig.~\ref{fig:QA_henneberg}, Poincaré plots are shown in Fig.~\ref{fig:henneberg_poincare} and \textcolor{black}{the self-consistent rotational transform profile is shown in Fig.~\ref{fig:henneberg_iota}. This dipole array solution exhibits the highest solution errors of the present work but still leaves the rotational transform essentially unchanged.} 

The primary peak field errors are on the inboard side, where the dipole coils cannot contribute much magnetic field strength. We compute a QFM and find that the two-term average quasisymmetry is $f_\text{QS} \approx 1.3 \times 10^{-3}$, improving slightly over the original coil solution~\cite{schuett2024exploring} but still reduced from the plasma design without coils that exhibits $f_\text{QS} \approx 1.5 \times 10^{-4}$.
Slightly less accurate coil solutions without dipole coils require $\approx 95.5$m of total coil length with two TF coils and $\approx 110.6$m of total coil length with three TF coils; the total length increases from two to three TF coils since the minimum coil-plasma and coil-coil distance must still be satisfied. It is challenging to fit more than a few TF coils on the inboard side of this compact stellarator. Despite the substantial reduction in the lengths of the TF coils, the total amount of superconductor required for all the coils is higher by approximately \textcolor{black}{29\%} ($30$ km) in the dipole coil configuration compared to the roughly equivalent two TF coil solution without dipoles. This is partly the case because the forces are high enough that around $200$ turns of wire are needed for almost every dipole coil. Whether the increased costs of purchasing more superconductor is a worthy trade-off for potential reduction in costs in coil manufacturing, assembly, and so forth is not addressed in this work. 
\textcolor{black}{Finally, note that there is substantial space from the top and bottom of the device for heating and diagnostic systems. For instance, there are multiple locations where a square-shaped port of roughly size $2$m $\times$ $2$m could be placed.}

\section{Discussion and Conclusion}\label{sec:conclusion}
We have implemented and tested several new coil objectives in the open-source SIMSOPT code for performing stellarator optimization. These new methods were used to provide the first demonstration of the reactor-scale optimization of stellarators with arrays of dipole coils. As we have shown, direct minimization of coil-coil forces and torques is often paramount for feasibility. Without using the spatial degrees of freedom of the coils, the required currents in the dipole coils will typically produce prohibitively large forces or the TF coil lengths may need to be significantly increased.
However, allowing the dipoles to move in space increases the geometric complexity of any support structure. Probably a balance between these options is the most feasible. For instance, often the dipole coils remain within $2$m of the plasma surface (and this can be enforced during optimization), so all center points roughly lie in a toroidal shell between $1.5-2$m from the plasma.

In practice, some hyperparameter tuning is needed for optimizing dipole arrays. For instance, often the benefits of the dipole coils, e.g. lower errors on the plasma surface, become quite modest after forces and torques on the dipole coils are minimized to below material tolerances, especially when the dipoles are fixed in space and orientation. \textcolor{black}{Like most forms of stellarator coil optimization, the number of coils must be specified before optimization, so in practice multiple optimizations are performed with varying numbers of dipole coils to find an optimal tradeoff between geometric complexity (and large forces) and solution accuracy on the plasma surface. We have also not considered that substantial space around the device will be required for heating and diagnostic systems. Such spatial constraints can easily be added to the optimization problem but a suitably accurate dipole solution will become harder to find.} These empirical facts raises the question whether certain stellarators really benefit from such dipole arrays. Nonetheless, we find several plausible reactor-scale coil solutions for which the dipole coils substantially simplify the TF coils by reducing the number, complexity, and total length. \textcolor{black}{In particular, the solution for the Schuett-Henneberg QA device exhibits only 2 unique TF coils and 16 unique dipole coils on the outboard side. Furthermore, it achieves solution errors low enough to retain good quasi-symmetry and leaves large regions of space for heating and diagnostic systems.}

More advantageous coil configurations, with or without dipole arrays, may be found in future work, as both problems exhibit numerous local minima and the tuning of the objective weights will depend on the tradeoff preferences of the user. For instance, we find that the solution can be sensitive to the initialization of the dipole array and TF coils; if the positions are fixed, there may be certain coils that are very challenging to find solutions with tolerable peak forces. If the coil locations and orientations are not fixed, there are many local minima and weights have to be tuned to avoid coil-interlinking and other unwanted behavior. During this work, we briefly investigated letting the dipole array coils change their shapes but this approach tended to suffer even further from local minima. 

Interestingly, we also find that net torques, despite their rough scaling as $I_i^2$, can often be very effectively reduced to negligible values by direct minimization and sufficient slack in the coil length objective. Future work includes the implementation of passive superconducting coil arrays, large hyperparameter and objective tradeoff scans, and designs for other stellarators.
To facilitate that future work, code speedups are required and GPUs should be utilized, as full-scale optimizations with $\sim 30-40$ unique dipole coils currently take a few hours to run on a Mac M2 Max CPU.  

\section{Acknowledgements}
This work was supported through a grant from the Simons Foundation under award 560651. The authors thank Mike Zarnstorff and Allen Boozer for some discussions on dipole coil arrays. The authors also thank Tobias Schuett and Sophia Henneberg for sharing their stellarator designs. 

\appendix
\section{Additional Pareto scans}\label{sec:appendix_pareto}
\begin{figure*}
    \centering
\includegraphics[width=\linewidth]{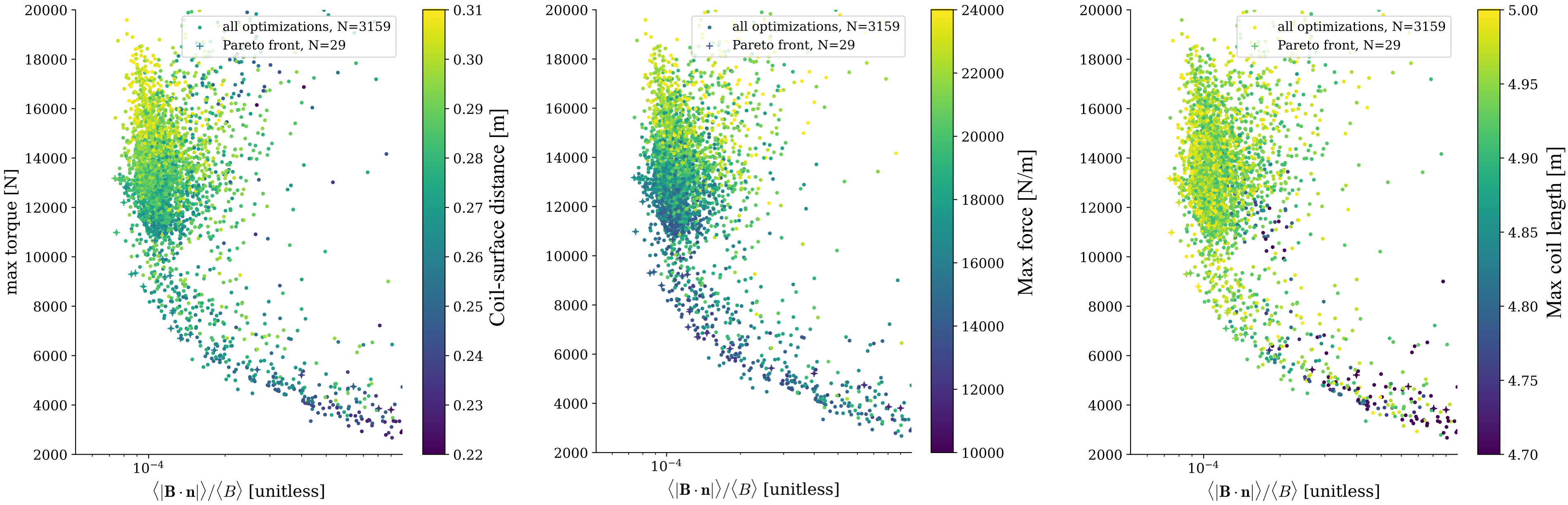}
    \caption{Significant correlation trends are observed from several thousand pointwise torque optimizations. Larger minimum coil-surface distance, larger coil forces, and larger coil lengths produce larger maximum torques.}
    \label{fig:torque_trends}
\end{figure*}

For completeness, we now present several thousand cold-start optimization runs in which only one of the force or torque objective terms, or the total vacuum magnetic energy, is optimized. The pointwise force results were already presented in Sec.~\ref{sec:preliminary_scans}. Fig.~\ref{fig:torque_trends} shows the coil quantities that exhibited a visible correlation when plotted against the pointwise torques and the normalized field errors. To a large extent, the trends when optimizing pointwise torques mirror the pointwise force trends. For instance, the pointwise torques tend to increase with increasing coil-surface distances, increasing pointwise forces, and increasing coil length. 

\begin{figure*}
    \centering
\includegraphics[width=\linewidth]{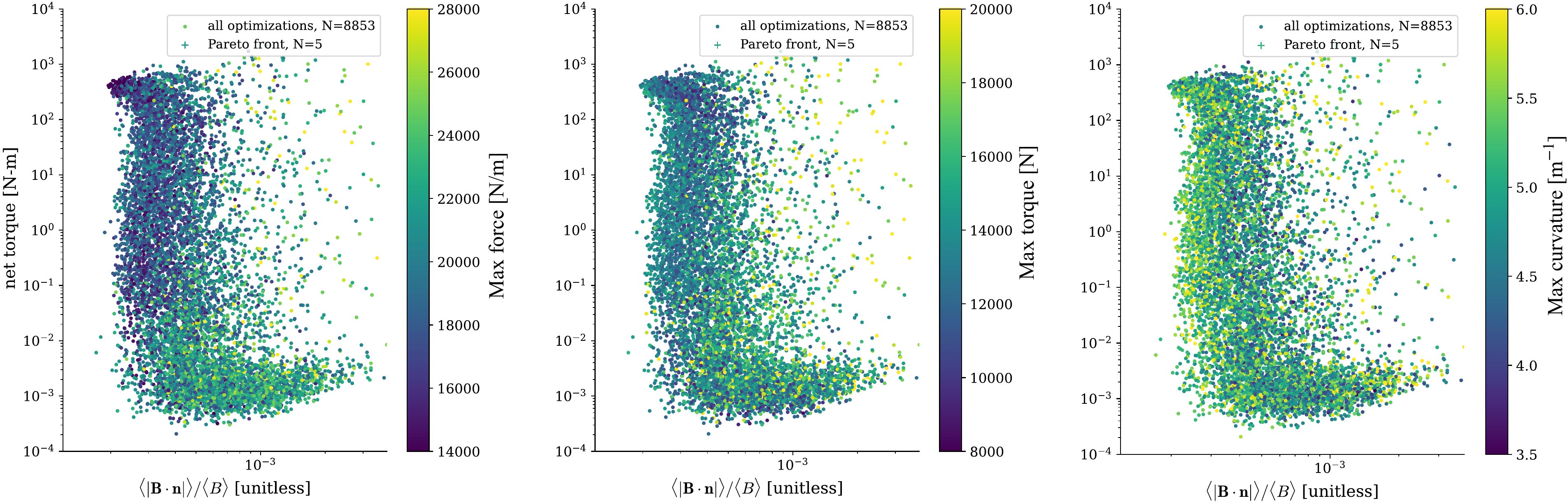}
    \caption{Clear trends are not observed from several thousand net torque optimizations. Net torques can be reduced to approximately $10^{-1}$ before a rough correlation develops between larger pointwise forces and torques and lower net torques. Most of the coil quantities exhibit variations like the maximum curvature plot on the right; larger maximum curvature often produces the smallest field errors, but the trend is essentially independent of the net torque.}
    \label{fig:nettorque_trends}
\end{figure*}

\begin{figure*}
    \centering
\includegraphics[width=\linewidth]{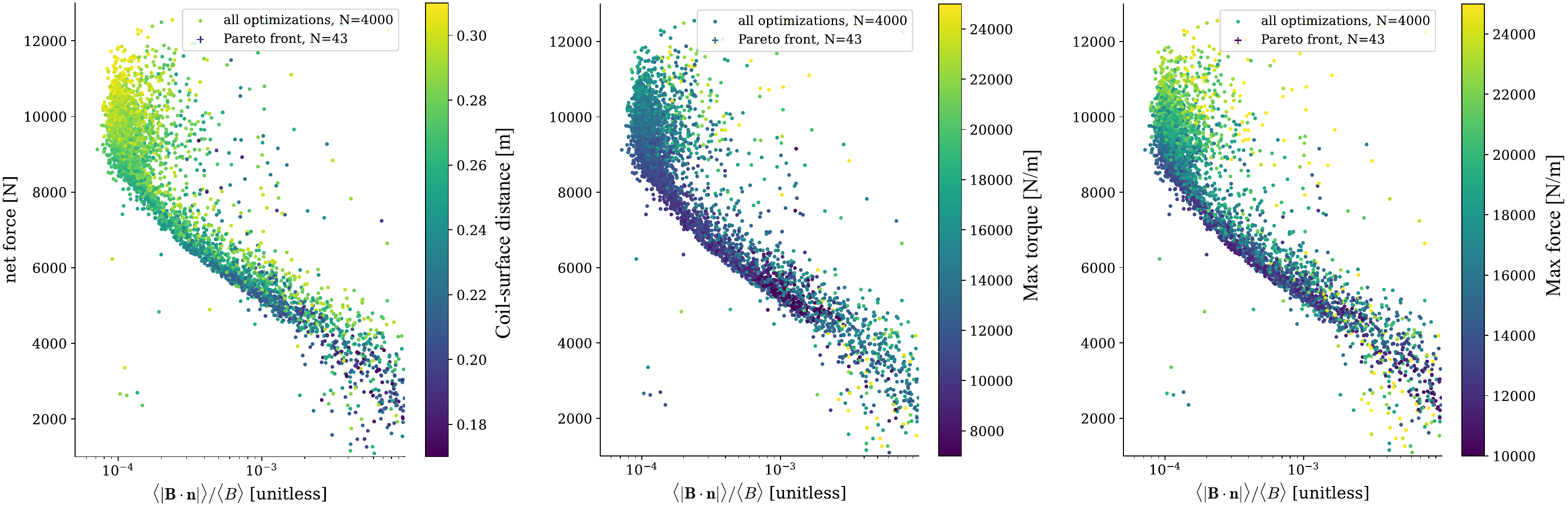}
    \caption{Significant correlation trends are observed from several thousand net force optimizations. Larger minimum coil-surface distance, larger pointwise torques, and larger pointwise forces tend to produce larger net forces.}
    \label{fig:netforce_trends}
\end{figure*}

Figures~\ref{fig:nettorque_trends} and~\ref{fig:netforce_trends} show similar illustrations for net torques and net forces. None of the illustrations for the net torques show particularly clear trends. For example, the net torque value appears to be uncorrelated even with the maximum pointwise force or the maximum pointwise torque until the net torque approaches negligible values of $10^{-1}$ N-m. This result supports the earlier finding in the present work that the net torque tends to be fairly uncorrelated with the other coil optimization objectives, and can be minimized without significant tradeoffs. In contrast, the net forces show substantial correlations; net forces increases at larger coil-surface distances, larger maximum pointwise torques, and large maximum pointwise forces, as might be expected. However, there are some low-torque, low-force solutions along the Pareto front that stay low as the net force decreases.

\begin{figure*}
    \centering
\includegraphics[width=\linewidth]{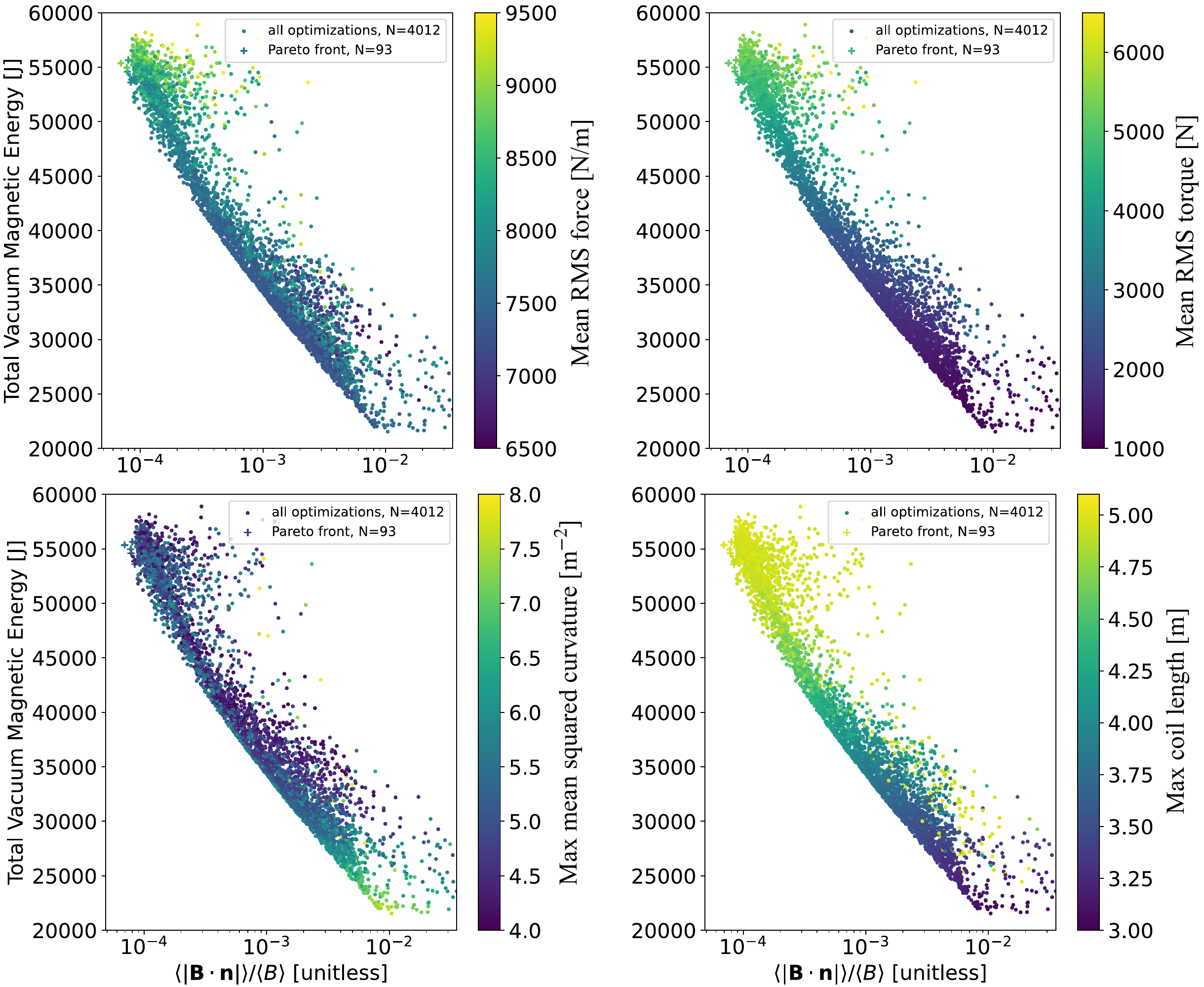}
    \caption{Significant correlation trends are observed from several thousand  optimizations of the total vacuum magnetic energy. Lower total energy correlates very strongly with reduced maximum coil lengths and reduced average root-mean-square torques. It also correlates well with reduced average root-mean-square forces and increased maximum mean-square curvature.}
    \label{fig:tve_trends}
\end{figure*}

Lastly, Fig.~\ref{fig:tve_trends} illustrates the correlations found for several thousand cold-start runs that minimize the total vacuum magnetic energy. The illustration shows an almost monotonic direct relationship between the total vacuum magnetic energy and the mean RMS torque and the max coil length. The correlations between these quantities are substantially stronger than any other illustrated trend and mirror the results in Guinchard et al. ~\cite{guinchard2024including} which found a similarly strong correlation between the total energy and the total coil length. We also see that minimizing the total energy tends to reduce the mean root-mean-square forces on the coils and increase the maximum mean-square curvature across the coils. There were several other quantities which showed some correlations, including coil-coil distances, coil-surface distances, and maximum coil curvature. Interestingly, the \textcolor{black}{maximum coil length} tended to decrease at lower total energy, even as the maximum mean-square curvature substantially increased. We neglect illustrating the coil-coil and coil-surface distance correlations because they seem to primarily be driven by the strong relationship between the total vacuum energy and the maximum coil length. In other words, if the maximum coil length is decreased, the coil-surface distance will generally decrease, and the coil-coil distance will increase.

\bibliography{DAO}
\end{document}